\documentclass[%
 reprint,
superscriptaddress,
nofootinbib,
 amsmath,amssymb,
 aps,
]{revtex4-2}

\usepackage{color}
\usepackage{amsmath}
\usepackage{amssymb}
\usepackage{graphicx}%
\usepackage{dcolumn}%
\usepackage{bm}%
\usepackage{hyperref}%
\usepackage{cleveref}
\usepackage{pbox}
\usepackage[utf8]{inputenc}
\usepackage[]{times}
\usepackage{ dsfont }

\usepackage{xcolor}
\definecolor{orange}{rgb}{1.0, 0.5, 0.0}

\usepackage{xcolor}
\definecolor{orange}{rgb}{1.0, 0.5, 0.0}

\begin{document}

\interfootnotelinepenalty=10000

\title{Impact of a mean field dynamo on neutron star mergers leading to magnetar remnants}

\author{Elias R. Most}
\email{emost@caltech.edu}
\affiliation{TAPIR, Mailcode 350-17, California Institute of Technology, 1200 E California Blvd, Pasadena, CA 91125, USA}

\date{\today}

\begin{abstract}
We investigate the impact of a mean field model for the $\alpha\Omega$-dynamo potentially active in the post-merger
phase of a binary neutron star coalescence.  We do so by deriving equations for ideal general relativistic magnetohydrodynamics (GRMHD) with an additional $\alpha-$term, which closely resemble their Newtonian counterpart, but remain compatible with standard numerical relativity simulations.  
We propose a heuristic dynamo closure relation for the
magnetorotational instability-driven turbulent dynamo in the outer layers
of a differentially rotating magnetar remnant and its accretion disk.
As a first demonstration, we apply this framework to the early stages of post-merger evolution ($\lesssim 50\, \rm ms$). We demonstrate that depending on the efficacy of
the dynamo action, magnetically driven outflows can be present with
their amount of baryon loading correlating with the magnetic field amplification. These
outflows can also contain precursor flaring episodes before settling into a
quasi-steady state. For the dynamo parameters explored in this work, we observe electromagnetic energy fluxes of up to $10^{50}\, \rm erg/s$, although larger amplification parameters will likely lead to stronger fluxes. 
Our results are consistent with the expectation that substantial dynamo amplification (either during or after the merger) may be necessary
for neutron-star remnants to power short gamma-ray bursts or precursors thereof.
\end{abstract}

\maketitle

\section{Introduction}

Neutron stars not only feature some of the most extreme conditions for
nuclear matter but can also contain some of the strongest magnetic fields
in the universe \cite{Baiotti:2016qnr}. 
In the context of neutron star mergers, magnetic fields
are crucial for driving electromagnetic counterparts such as kilonova
afterglows \cite{Metzger:2019zeh} and gamma-ray bursts \cite{Ruiz:2021gsv},
see also GRB170817A for a short-duration gamma-ray burst jointly detected with a gravitational wave event \cite{LIGOScientific:2017ync}.

Based on long-term simulations of neutron star -- black hole coalescence \cite{Gottlieb:2023est,Hayashi:2021oxy,Hayashi:2022cdq}, it has recently been suggested that while long gamma-ray bursts may be powered by black hole disk systems, short gamma-ray bursts may require the presence of a magnetar engine \cite{Gottlieb:2023sja}. However, the precise conditions and timescales probed in the simulations underlying these conclusions may sensitively depend on the magnetic field geometry produced in the collisions \cite{Etienne:2011ea,Etienne:2012te,Ruiz:2020via,Most:2021ytn,Hayashi:2022cdq}, with additional uncertainties coming from pre-merger fields \cite{Paschalidis:2014qra,Ruiz:2020via} (see also Refs. \cite{Paschalidis:2013jsa,East:2021spd,Carrasco:2021jja,Most:2023unc}).

Furthermore, the presence of a blue component in the electromagnetic afterglow of GRB170817A
appears to require additional magnetic and neutrino-driven mass ejection, likely to
originate from the surface of a hot hypermassive neutron star (HMNS) formed
in the merger \cite{Metzger:2014ila,Fujibayashi:2017puw,Fahlman:2018llv,Metzger:2018qfl,Kawaguchi:2022bub}. Recent numerical relativity simulations of this
process seem to support this conclusion \cite{Mosta:2020hlh,Curtis:2023zfo,Combi:2023yav}. 

However, our understanding of the dynamics and evolution of the magnetic
field in such a magnetar merger remnant remains incomplete. This is largely a
result of underresolved dynamo physics during the merger \cite{Price:2006fi,Kiuchi:2015sga} and
inside the post-merger remnant \cite{Kiuchi:2017zzg,Aguilera-Miret:2021fre}. 
Although small-scale dynamo processes at merger seem to be driven by Kelvin-Helmholtz instabilities at
the shearing interface layer between the two stars \cite{Kiuchi:2015sga}, the effect of
macroscopic amplification in the bulk of the star remains uncertain
\cite{Aguilera-Miret:2023qih}.  Similarly, additional processes such as Rayleigh-Taylor
instabilities on stellar surfaces may contribute to the overall
amplification \cite{Skoutnev:2021chg}. Recently, very high-resolution numerical relativity simulations of the post-merger remnant have been reported to show the presence of a magnetorotational instability \cite{Balbus:1991ay} (MRI)-driven $\alpha\Omega-$ dynamo \cite{Kiuchi:2023obe}, which may also depend on the chemical stratification present in the remnant \cite{Radice:2023zlw}, see also \cite{Guilet:2016sqd}. Similar processes may also operate in the accretion disk \cite{Liska:2018btr,Christie:2019lim,Jacquemin-Ide:2023qrj}.
Clarifying the precise impact of such under-resolved processes on the neutron star remnant will be central to our understanding of whether magnetar engines formed in mergers can power short gamma-ray bursts. This is particularly relevant, as the neutrino-driven stellar winds may strongly baryon load the outflows, potentially lowering the luminosity \cite{Dessart:2008zd,Fujibayashi:2017puw,Metzger:2018qfl}.

Incorporating under-resolved dynamo physics into neutron star mergers is challenging.
Previous effective numerical models have resorted to simple exponential modifications of the evolution equations \cite{Giacomazzo:2014qba,Palenzuela:2015dqa}, or have made use of effective Large-Eddy approaches \cite{Aguilera-Miret:2020dhz,Palenzuela:2021gdo}. Others, inspired by studies of accretion disk dynamos \cite{Bucciantini:2012sm,Sadowski:2014awa,Tomei:2019zpj}, have used effective mean field models \cite{Shibata:2021bbj}.  Here, we follow the latter approach but focus mainly on the ability of launching winds and jet-like outflows from the surface of the hypermassive neutron star. Due to the natural shearing background present in the differentially rotating remnant star \cite{Hanauske:2016gia,Kastaun:2016yaf}, one possibly active mechanism for mean field amplification is the $\alpha\Omega-$dynamo \cite{Parker:1958zz,1978mfge.book.....M}.
Recent works have shown that the presence of such an effect inside the outer layers of the neutron star can lead to magnetic field breakout and the launching of (intermittent) jet-like outflows \cite{Most:2023sft,Kiuchi:2023obe}.

Effective models including mean field $\alpha$-terms in numerical simulations have been developed by
various communities for applications ranging from early universe cosmology
\cite{Brandenburg:2017rcb}, galaxy dynamics \cite{Rieder:2017qes}, to supernovae
\cite{White:2021erz} and compact objects \cite{Bucciantini:2012sm,Tomei:2019zpj,Shibata:2021bbj}.  
While most of these formulations have considered Newtonian magnetohydrodynamics (MHD), in which the inclusion of an effective
electromotive force can be straightforward, the relativistic
context poses extra challenges. 
Previous approaches have chosen to
include effective $\alpha$- and Hall terms in a dynamical Ohm's law \cite{Bucciantini:2012sm,Shibata:2021bbj}.  
In the relativistic context, this requires the solution of an
additional evolution equation for the electric field, which becomes
numerically stiff in the perfectly conducting limit, and requires implicit
time stepping to be handled stably \cite{Palenzuela:2008sf,Ripperda:2019lsi}. 
On the other hand, these approaches have been demonstrated to have no well-posed initial value
problem \cite{Schoepe:2017cvt}. Although from a physics point of view full dissipative approaches might be more desirable \cite{Andersson:2021kfk,Most:2021uck}, the
additional computational overhead and the loss of robustness established for GRMHD simulation \cite{Kalinani:2021ofv}, makes this less appealing, especially when full nonideal effects are not required for the numerical modeling.\\

Complementing these works, we here present a new formulation to
systematically incorporate the $\alpha-$ effect into numerical general relativistic MHD
simulations. Together with the shear flow naturally present in the
post-merger system, this naturally enables a subgrid model for the
$\alpha\Omega$-dynamo. 
While the formulation is generic, we apply it here specifically to the case of a post-merger
magnetar remnant.

Our paper is structured as follows.
In Sec. \ref{sec:methods} we present the general approach to modeling the
$\alpha-$ effect, before introducing a generalization to the
relativistic regime. We then discuss a heuristic subgrid model, before
outlining the particular steps needed to
incorporate them into a numerical relativity code.
In Sec. \ref{sec:results},  we present an initial assessment of this model in the
early post-merger phase of a magnetar remnant.

\section{Methods}\label{sec:methods}
In this work, we study the impact of a mean field dynamo model on
the evolution of a magnetar remnant formed in a neutron star collision.
To this end, we have implemented a new approach to incorporating
subgrid dynamo models, in particular, the $\alpha-$effect into 
non-resistive general-relativistic magnetohydrodynamics (GRMHD) simulations.
In the following, we will first describe a formulation of the
$\alpha-$effect in Sec. \ref{sec:alpha}, before providing details on the
numerical implementation and simulation setup in Sec. \ref{sec:numerics}.

As the main theoretical framework for modeling the evolution of stellar
matter, we solve the Einstein-Maxwell-Fluid system in a
magnetohydrodynamic approximation \cite{Duez:2005sf}. More specifically, we solve
\begin{align}
    G^{\mu\nu} = 8\pi T^{\mu\nu}\,, \label{eqn:einstein}\\
    \nabla_\mu T^{\mu\nu} = 0\,, \label{eqn:hydro} \\
    \nabla_\mu \,^{\ast}\! F^{\mu\nu} =0\,, \label{eqn:maxwell}
\end{align}
where $G_{\mu\nu}$ is the Einstein tensor, 
$F^{\mu\nu}$, and , $\,^{\ast}\!F^{\mu\nu}$ are the Maxwell tensor and its dual.  
Introducing the fluid four-velocity $u^\mu$, we can decompose the field strength tensor into
comoving electric, $e^\mu$, and magnetic, $b^\mu$, fields \cite{Palenzuela:2012my},
\begin{align}
    F^{\mu\nu} &= u^\mu e^\nu - e^\mu u^\nu + \varepsilon^{\mu\nu\kappa\lambda} b_\kappa u_\lambda\,, \label{eqn:F}\\
    \,^{\ast}\!F^{\mu\nu} &= u^\mu b^\nu - b^\mu u^\nu - \varepsilon^{\mu\nu\kappa\lambda} e_\kappa u_\lambda\,. \label{eqn:Fstar}
\end{align}
Here $\varepsilon^{\mu\nu\kappa\lambda}$ is the four-dimensional Levi-Civita tensor.
We further assume that the plasma dynamics are governed by a perfect fluid
described by its rest-mass density, $\rho$, total energy density
$\epsilon$, and pressure $P$, in addition to its fluid four-velocity,
$u^\mu$.  The combined system is described by the total energy-momentum
tensor 
\begin{align}
    T^{\mu\nu} = &\left[\epsilon+P +e^2 + b^2 \right] u^\mu u^\nu + \left[P + \frac{1}{2}\left( e^2 + b^2 \right) \right] g^{\mu\nu} \nonumber \\
    &- b^\mu b^\nu - e^\mu e^\nu  + e_\alpha b_\beta u_\gamma \left[ u^\mu \varepsilon^{\nu \alpha \beta \gamma} + u^\nu \varepsilon^{\mu \alpha \beta \gamma} \right]\,. \label{eqn:Tmunu}
\end{align}

In order to evolve the system \eqref{eqn:einstein}-\eqref{eqn:maxwell}, we
need to specify closures for the pressure and electric field. For a non-resistive plasma, we
need not evolve the electric field independently. In other
words, we need to provide an equation of state, $P=P\left(\rho, \epsilon,
\ldots\right)$, and a relation $e^\mu = e^\mu\left(u^\mu, b^\mu,\ldots
\right)$ \footnote{We point out that alternatively, in a resistive
approach, we could also solve an evolution equation for the electric field
$E^i$, which would couple to the fluid via an Ohm's law \cite{Palenzuela:2008sf,Bucciantini:2012sm,Dionysopoulou:2012zv,Qian:2016lyn,Ripperda:2019lsi} However, here we
consider the perfectly conducting ideal-MHD limit appropriate for neutron
star mergers \cite{Harutyunyan:2018mpe}.}.

\subsection{Subgrid dynamo physics}\label{sec:alpha}

We now want to provide a brief summary of mean field dynamo
theory. For further background, see, e.g., Ref. \cite{Brandenburg:2004jv}
for a review. To better illustrate the approach taken here, we
first review a Newtonian formulation of the problem, before doing the same
in a fully covariant, general-relativistic context.

Consider the flat space evolution equation for the magnetic field $B^i$, 
\begin{align}
    \partial_t { B^i} + \varepsilon^{ijk} \partial_j  E_k =0\,,
\end{align}
where $\varepsilon^{ijk}$ is the Levi-Civita tensor. Consistent with the
high conductivity found inside neutron stars, we further assume the ideal
MHD limit $E_i = - \varepsilon_{ijk}v^j B^k$, where $v^k$ is the 3-fluid
velocity \cite{sturrock1994plasma}.
The main reasoning behind dynamo mean field theory is to capture the effect
of (small-scale) fluctuations operating on top of a background mean field.
To this end, one may decompose the magnetic and velocity fields  
\begin{align}
    B^i &= \bar{B}^i + \delta B^i\,,\\
    v^i &= \bar{v}^i + \delta v^i\,,
\end{align}
into an averaged mean field $\bar{B^i},\bar{v}^i$, and small fluctuations
$\delta{B^i},\delta{v}^i$, such that on average $\left<B^i\right> = \bar{B}^i$ and
$\left<v^i\right> = \bar{v}^i$, respectively.  When evaluating the effective mean electric field, in the same way, one finds that
\begin{align}
\bar{E}^i = - \varepsilon^{ijk}\bar{v}_j \bar{B}_k
- \left<\delta \mathbf{v} \times \delta \mathbf{B}\right>^i.
\end{align}
Although individual fluctuations vanish, their second-order correlation does
not. In mean field dynamo theory, this system is now closed by performing a
first-order expansion of these fluctuations in terms of the mean
magnetic field \cite{Gruzinov:1994zz},
\begin{align}
\left<\delta \mathbf{v} \times \delta \mathbf{B}\right>^i &\approx
 \alpha^i_j \bar{B}^j  - \beta^i_l \varepsilon^{ljk} \partial_j \bar{B}_k\,, \\
&\approx \alpha^i_j \bar{B}^j - \beta^i_k J^k\,,
\label{eqn:correlation_newtonian}
\end{align}
where we have introduced tensorial mean field dynamo coefficients
$\alpha^i_j$ and $\beta^i_j$. Indeed, for certain processes such as
magnetorotational instability-driven dynamos, can be computed from
local simulations \cite{Gressel:2015mxa}. In the last line, we have used that in the small velocity limit
$\nabla \times \mathbf{B} \approx \mathbf{J}$, where $\mathbf{J}$ is the electric
current.  In summary, 
\begin{align}
E^i = - \varepsilon^{ijk}\bar{v}_j \bar{B}_k
-\alpha^i_j \bar{B}^j + \beta^i_j J^j\,, \label{eqn:E_newtonian}
\end{align}
From Eq. \eqref{eqn:E_newtonian} it is now apparent that $\beta$ acts as
an effective anisotropic resistivity.\\

We now want to generalize the forgoing discussion to the relativistic
context.
In doing so, we effectively have two options. We could consider
fully covariant first-order perturbation theory \cite{Armas:2022wvb}, or instead choose
to generalize the Newtonian expressions \eqref{eqn:correlation_newtonian} and \eqref{eqn:E_newtonian}, which we pursue. While our approach is similar to the
one put forward by Ref. \cite{Bucciantini:2012sm} (see also Ref.
\cite{Shibata:2021bbj}), we will make further simplifications concerning
the resistive timescales of the plasma. These will allow us to obtain
equations that do not suffer from stiffness problems in the ideal MHD limit
and do not require implicit time stepping
\cite{pareschi_2005_ier}.
Our discussion begins with a dynamical Ohm's law inspired by relativistic
14-moment closure theory \cite{Most:2021uck}, which we have supplemented
with tensorial mean field dynamo terms $\alpha^\mu_\nu$ and $\beta^\mu_\nu$,
\begin{align}
  \tau u^\nu \nabla_\nu J^{<\mu >} = - J^\mu +  \eta^{-1} \bar{e}^\mu
  + \alpha^\mu_\nu \bar{b}^\nu  - \beta^\mu_\nu J^\nu\,,
  \label{eqn:Ohms}
\end{align}
where $J^\mu$ is the electric current, $\eta$ the electric resistivity,
$\bar{e}^\mu$ is the mean
electric field and $\bar{b}^\mu$. We have also introduced a collisionality timescale $\tau$. These quantities are by definition projected into the fluid frame by means of the mean four-velocity of the fluid $\bar{u}^\mu$. This implies that $u_\mu \alpha^\mu_\nu = \beta^\mu_\nu u_\mu =0$. 
In the following, we shall drop any explicit notation of
mean field quantities for improved readability. 
We now make several assumptions, applicable to the
case of neutron star merger dynamics. Firstly, we assume that the collision
timescale, $\tau$, associated with the mean-free path of the system, is
much shorter than large-scale variations of the system \cite{Harutyunyan:2018mpe}. 
In this limit, we may neglect the term on the left-hand side in Eq.
\eqref{eqn:Ohms}. We caution, however, that without further simplifications
concerning the resistivity, $\eta$, the system would likely lose strong
hyperbolicity \cite{Schoepe:2017cvt}, though in practice numerical solutions of this system can still be found \cite{Palenzuela:2008sf,Bucciantini:2012sm,Qian:2016lyn,Ripperda:2019lsi}. Alleviating this fact, and in line with the almost
perfect conductivity encountered in neutron star matter \cite{Harutyunyan:2018mpe}, we will
now reduce the equations to their non-resistive limit, $\eta
\rightarrow 0$. To this end, we re-write 
\begin{align}
  J^\mu = \left(\frac{\eta^{-1}}{\mathds{1} +\beta}\right)^\mu_\nu e^\nu +
         \tilde{\alpha}^\mu_\nu b^\nu\,,
\end{align}
where $\tilde{\alpha}^\mu_\nu = \left(\mathds{1}+ \beta\right)^{-1}\,^\mu_\lambda
\alpha^\lambda_\nu$. We can now see that the $\beta$-effect creates an
anisotropic conductivity (similar to the Hall effect \cite{sturrock1994plasma}). Indeed, such
effects have been studied in the context of shear-current dynamo amplification
\cite{Squire:2015jma}. In this work, we do not consider such effects and will only study
the limit where the mean field dynamo operates on timescales much longer than the
effective resistive timescale. 
In proceeding along these lines, we now make the crucial assumption that
dynamo effects will always grow relative to the resistive timescale. In
other words, $ \kappa := -\eta
\left|\alpha\right|$ does not vanish in the limit of zero resistivity. 
This is a well-justified assumption if we consider
modeling an ultimately kinematic dynamo, which will operate on viscous scales larger than the resistive scale \cite{Guilet:2016sqd}. This allows us to
re-express the comoving electric field as,
\begin{align}
    e^\mu = \kappa^{\mu}_{\nu}\, b^\nu\,, \label{eqn:dynamo_final}
\end{align}
where we have re-expressed $\kappa^\mu_\nu :=  -\eta \alpha^\mu_\nu$ .
Eq. \eqref{eqn:dynamo_final} is the constitutive closure relation of our mean field
dynamo model.
This lets us draw an important conclusion, for the approximations we make.
Since we assume that we are essentially always nonresistive, any nonideal
dynamo growth of the magnetic field must come from a nonzero $\alpha$-term, i.e. nonvanishing components of $\kappa^\mu_\nu$. As we will see later on,
making this assumption allows us to use an ideal GRMHD code, with only
minor modifications.

\subsection{3+1 mean field dynamo equations}
Having outlined the form of the above equations, we next proceed to recast them into a form suitable for numerical integration.  This is
done in two steps. First, we will show that the dynamo action
\eqref{eqn:dynamo_final} does not require us to evolve the electric field
itself, but rather will lead to ideal-MHD-like closure relations.
In the second step, we highlight modifications to the GRMHD equations
necessary to include these effects. In the following, we adopt the ADM
split of spacetime \cite{Arnowitt:1962hi},
\begin{align}
  {\rm d}s^2 = \left(- \alpha^2 + \beta_k \beta^k\right) {\rm d}t^2 + 2
  \beta_j {\rm d}x^j {\rm d} t + \gamma_{\ij} {\rm d}x^i {\rm d} x^j\,,
  \label{eqn:metric_adm}
\end{align}
where $\alpha$ is the lapse function, $\beta^i$ the shift vector, and
$\gamma_{ij}$ the spatial metric.

While our dynamo closure \eqref{eqn:dynamo_final} links comoving electric and magnetic fields, we need to recover the electric, $E^\mu = n_\nu F^{\mu\nu}$, and magnetic, $B^\mu = n_\nu \,^{\ast}\!F^{\mu\nu}$, fields as seen by a Eulerian observer corresponding to the space-time normal $n_\mu = \left(-\alpha, 0,0,0\right)$. Using these definitions, we can compute
\begin{align}
    B^\mu =  \Gamma b^\mu - u_i B^i u^\mu +\varepsilon_{(3)}^{\mu\lambda\nu} u_\lambda \kappa_{\nu \alpha} b^\alpha\,,
\end{align}
where we have used that $\alpha b^0 = u_i  B^i$, as well as the dynamo
relation \eqref{eqn:dynamo_final}. We have also introduced the
three-dimensional Levi-Civita tensor embedded in a four-dimensional
space, $\varepsilon_{(3)}^{\mu\nu\lambda} = n_\kappa
\varepsilon^{\mu\kappa\nu\lambda}$, as well as the Lorentz factor, $\Gamma
= - n_\mu u^\mu$. It is also convenient to introduce the spatial velocity
$v_\mu = \left(u_\mu - \Gamma n_\mu\right) /\Gamma$. Next, we need to solve this equation for $b^\mu$, so
that we can express the comoving magnetic field in terms of $B^i$,
\begin{align}
    b^\alpha &= \frac{1}{\Gamma \left( \delta^\mu_\alpha + \varepsilon_{(3)}^{\mu\lambda\nu} v_\lambda \kappa_{\nu \alpha} \right)} \Delta^{\mu\nu} B_\nu\,,\\
    &= b^\alpha_{\rm ideal} - \varepsilon_{(3)}^{\alpha\lambda\nu} v_\lambda \kappa_{\nu \mu} b^\mu_{\rm ideal} + \mathcal{O}\left(\left|\kappa\right|^2\right)\,, \label{eqn:b_series}
\end{align}
where $b^\mu_{\rm ideal} = \Delta^{\mu\nu} B_\nu /\Gamma$ is the comoving
magnetic field in ideal (non-resistive) GRMHD \cite{Duez:2005sf}, and $\Delta^{\mu\nu}
= g^{\mu\nu} + u^\mu u^\nu$ is the fluid-frame projector.  We point out
that as a geometric series, Eq.\eqref{eqn:b_series} will converge as long
as the dynamo timescales are much longer than the resistive timescale, i.e.,
$\left|\kappa\right| \ll 1$. 
Furthermore, the expansion in $\left|\kappa\right|$ fundamentally preserves covariance of the equations, but will make the solution inaccurate should the dynamo begin to act on timescales comparable to the resistive scale. 

\subsubsection{A pseudo-scalar dynamo model}

While modeling specific dynamo processes, such as those associated with the
MRI, will require anisotropic dynamo
coefficients \cite{Gressel:2015mxa,Hogg:2018zon},
incorporating
anisotropic dynamo terms will require more substantial modifications to existing GRMHD
codes, similar to using a fully resistive
\cite{Palenzuela:2008sf,Bucciantini:2012sm,Ripperda:2019lsi,Shibata:2021bbj}, or a dissipative MHD code \cite{Most:2021rhr}. Circumventing these challenges, in this first application of this model we will
restrict ourselves to a simplified dynamo model commonly employed in
cosmology \cite{Brandenburg:2017rcb}, supernova \cite{White:2021erz}, and relativistic contexts \cite{Tomei:2019zpj}. For a (pseudo-)scalar 
$\alpha-$effect, $\kappa^\mu_\nu = \kappa \Delta^\mu_\nu$, the equations simplify
considerably.\\
In the following, we, therefore, assume
\begin{align}
    e^\mu = \kappa b^\mu\,.
\end{align}
Using this definition, we can easily see that
\begin{align}
    F^{\mu\nu} = \kappa\,^\ast\! F^{\mu\nu} + \varepsilon^{\mu\nu\kappa\lambda} b_\kappa u_\lambda + \mathcal{O}\left(\kappa^2\right),
\end{align}
which leads to an electric field in the normal observer frame,
\begin{align}
        E^i =& - \varepsilon^{ijk}v_j B_k + \kappa \left[ \left(1- v^2\right) B^i  +  \left(v_l B^l\right) v^i \right] \nonumber \\ &+ \mathcal{O}\left(\kappa^2\right)\,,
    \label{eqn:Efield_scalar}
\end{align}
which we have expanded to leading order in $\kappa$. This expression is remarkably similar to the Newtonian limit given in Eq. \eqref{eqn:E_newtonian} (in the absence of a $\beta$-term), which it recovers in the small velocity limit,
\begin{align}
        E^i =& - \varepsilon^{ijk}v_j B_k + \kappa B^i + \mathcal{O}\left(\kappa^2, \kappa v^2\right)\,.
    \label{eqn:Efield_scalar_Newton}
\end{align}
We point out that despite the apparent expansion in $v^2$ both expressions \eqref{eqn:Efield_scalar_Newton} and \eqref{eqn:Efield_scalar} are fully covariant, but correspond to either the full or only partial first-order contribution in $\kappa$ in Eq. \eqref{eqn:b_series}. 
For the background flows inside the hypermassive neutron star remnants $v\lesssim 0.3$, we likely do not expect a large difference between the two approximations. For completeness, we opt to use Eq. \eqref{eqn:Efield_scalar} and 
\begin{align}\label{eqn:b_comov_kappa}
b^i = b^i_{\rm ideal} - \frac{\kappa}{\Gamma}\varepsilon^{ijk} v_j B_k + \mathcal{O}\left(\kappa^2\right)\,.
\end{align}

Similarly, the stress-energy tensor results in
\begin{align}\label{eqn:Tmunu_ideal}
    T^{\mu\nu} = &\left[\epsilon+P + b^2 \right] u^\mu u^\nu + \left[P +
    \frac{1}{2}b^2 \right] g^{\mu\nu} 
    - b^\mu b^\nu \nonumber\\  &+\mathcal{O}\left(\kappa^2\right)\,,
\end{align}
which is consistent with ideal GRMHD. In the scalar limit, the comoving
electric and magnetic fields always line up, so that no effective heat flux
contributions are present. This is equivalent to stating that the presence of the scalar
dynamo terms does not induce a nonideal Poynting flux $S_i^{\rm EM}$, which gets
modified in the anisotropic case. \\

It is instructive to compare this formalism with previous approaches to
modeling scalar $\alpha$-dynamo effects in the GRMHD literature
\cite{Bucciantini:2012sm,Tomei:2019zpj,Shibata:2021bbj}. 
Essentially, the starting point for these formulations is a fully
resistive description, where the dynamo term is included as part of the
Ohms law \eqref{eqn:Ohms}. The advantage of such an approach is that the resistive
timescale can be explicitly fixed, allowing dynamo amplification (to all
orders in $\kappa$) to occur on fully controllable timescale, 
whereas in our approach $\kappa$
only fixes the ratio of the dynamo growth time to (effectively) numerical
resistivity. On the other hand, our approach allows us to only minimally
modify the ideal GRMHD equations, while the resistive equations become
stiff in the perfectly conducting limit and require special numerical
treatment \cite{Palenzuela:2008sf}. 

\subsection{Subgrid model parameters}\label{sec:subgrid}

Having described the method, the final step in completing this model is a
specification of the dynamo closure parameters. These will strongly depend
on the microphysics that is being modeled. In the case of relativistic accretion
disks, the capture of under-resolved MRI turbulence has been accomplished using
different prescriptions \cite{Sadowski:2014awa,Tomei:2019zpj}, which have been tested against full
three-dimensional resolved simulations, or have been inspired by local
shearing box calculations \cite{Gressel:2015mxa,Gressel:2022cxr}. Rather than relying on such
calculations, which may not directly translate to the HMNS case, we take values inspired by a recent very high-resolution
simulation \cite{Kiuchi:2023obe}. There it was found that the outer (MRI unstable) layers 
of the neutron-star remnant can drive a large-scale $\alpha\Omega-$dynamo
\cite{Kiuchi:2023obe}. These results further imply that
\begin{align}
    \kappa_{\rm HMNS} \simeq 0.025 - 0.035\,,
\end{align}
where we point out that $\kappa$ is a pseudo-scalar and has dimensions of a velocity, which we express relative to the speed of light. We caution that in our case $\kappa$ is defined with the opposite sign and also includes an effective resistivity contribution that we have absorbed into the above expression.
While $\kappa_{\rm HMNS}$ determines the
growth rate of the $\alpha\Omega-$dynamo, it does not restrict its
saturation level. Physically, saturation is expected to happen when the
relevant reservoir of (largely kinetic) energy is converted into
magnetic energy. In the absence of this self-limiting feedback on the
mean field dynamo term when $\kappa$ is prescribed, saturation has to be imposed manually using
different criteria.\\
In practice, we opt to limit
\begin{align}\label{eqn:kappa_max}
    \kappa = \kappa_{\rm HMNS} \max\left( 0, \Delta \left[\rho, b^2, T,\ldots\right] \right)\,,
\end{align}
where $\Delta$ is an indicator function for saturation, which is reached for $\Delta < 0$.
The formalism presented here will work for every choice of $\Delta$. For our specific application, we now suggest a phenomenological model for the saturation function $\Delta$ inspired by very high-resolution simulations \cite{Kiuchi:2017zzg}. When using several prescriptions of $\Delta$, we opt to use Eq. \eqref{eqn:kappa_max} jointly over all them.

\begin{enumerate}
    \item {\bf Magnetization-dependent saturation limit}
    
    Here we limit based on the strength of the generated radial magnetic field $B^r$. This is meaningful since the $\alpha\Omega-$ dynamo should generate a radial magnetic field only up to a fraction of the toroidal field, $B^\phi$. If $B^r \gtrsim B^\phi$ then the $\alpha$-effect would continue to generate more toroidal field via an $\alpha^2\Omega-$dynamo, which is not supported by global simulations \cite{Kiuchi:2023obe}.
    We in turn adopt,
    \begin{align}
        \Delta_{\alpha^2} = 1 - \frac{1}{\omega}\frac{B^r}{B^\phi}\,,
    \end{align}
    where $\omega<1$ is the fraction of radial magnetic field.

    \item {\bf Turbulent energy limit}
    
    In principle, the dynamo should be fueled by the available turbulent energy in the flow.
    Defining this number is not trivial and has been extensively investigated in the Newtonian large-eddy literature \cite{2011A&A...528A.106S,2022MNRAS.513.6028L,Miravet-Tenes:2022ebf,Miravet-Tenes:2023see}.
    The major challenge lies in defining what fraction of the turbulent kinetic energy is available to be converted into magnetic energy, as well as estimating the available turbulent kinetic energy in the first place.\\
    Assuming that the driving and decay (to smaller scales) of kinetic energy are in steady state, we can show that the saturation-level magnetization $\sigma_{\rm turb}$ should be determined by the turbulent kinetic energy.
    More specifically, we find that (see Appendix \ref{app:turb}),
    \begin{align}
        \sigma_{\rm turb} = \xi \ell_{\rm turb}^2 \sigma_{\mu\nu} \sigma^{\mu\nu}\,,
    \end{align}
    where $\ell_{\rm turb}$ is an effective mixing length, $\xi < 1$ the fraction of turbulent kinetic energy converted into magnetic energy, and $\sigma_{\mu\nu}$ is the shear stress tensor. 
    The dynamo should then saturate, when the turbulent energy budget has been used up, i.e., when
    \begin{align}
    \Delta_{\rm turb} = 1 - \frac{\sigma}{\sigma_{\rm turb}}\,,
\end{align}
    where $\sigma = b^2/\rho$ is the magnetization.
    Local fluctuations in these quantities sometimes lead to large spurious values in individual grid cells. As a safety measure, we decided to cap the reference magnetization accordingly, i.e., enforce $\sigma_{\rm turb} < \sigma_{\rm max}\ \sim 0.01$.\\

    \item {\bf Effective saturation closure }

    \noindent In practice, subgrid models may be useful in capturing the effective mixing length of MRI-driven turbulence.
    From a first-principles point of view, the turbulent mixing length should be limited by the largest wavelength, $\lambda_{\rm MRI}$, of the poloidal MRI \cite{Duez:2006qe},
    \begin{align}
        \ell_{\rm MRI} \simeq \lambda_{\rm MRI} \sim \frac{2\pi v_A^z}{\Omega}\,,
    \end{align}
    where $v_A^z$ is the Alfven speed in the poloidal direction. 
    However, this is not a robust criterion in low-resolution simulations, since $\ell_{\rm MRI} \sim \left|b^z \right|$. In other words, if $b^z$ is not fully developed
    in a low resolution simulations, the mean field feedback (that may contribute to its development) would not set in, 
    self-limiting the applicability of the mean field prescription.
    We therefore opt to not use a magnetic field-dependent sub-grid model.
    Based on very high-resolution simulations of a neutron-star merger remnant \cite{Kiuchi:2017zzg}, Ref. \cite{Radice:2020ids} has proposed a density-dependent fit formula for $\ell_{\rm MRI}$, which is given by
    \begin{align} \label{eqn:lmix}
        \ell_{\rm MRI}^{\rm HMNS}  &= \max \left(0, a x \, \exp \left[- \left|b x\right|^{5/2}\right]\right) \left[m \right] ~\,, \\
        x &= \log_{\rm 10} \left(\rho / \rho_\ast\right)\,, \nonumber
    \end{align}
    where $a=22.31984$, $b=-0.4253832$ and $\rho_\ast = 1.966769\times 10^9\, {\rm g cm^{-3}}$ \cite{Radice:2020ids}. 
    This prescription matches $\lambda_{\rm MRI}$ in a certain density regime but is independent of the local magnetic field in the simulation (see Fig. \ref{fig:lmix}). Therefore, we opt to use $\ell_{\rm MRI}^{\rm HMNS}$ as the fiducial sub-grid model in this work. \\

    {\bf Saturation limit:} Finally, we need to quantify the uncertainty factor $\xi$ within the model. This is necessary for a number of reasons. First, for algorithmic simplicity, we have assumed a steady state for the turbulent kinetic energy, which is likely not true and will depend on the effective scales probed by the
    models (in our case the numerical grid scale). Indeed, when considering
    large eddy simulations of post-merger magneto-turbulence, the kinetic energy and especially the amount of turbulence seems to depend on the numerical resolution \cite{Aguilera-Miret:2020dhz,Aguilera-Miret:2023qih}.\\ 
    Secondly, the assumption of the turbulent mixing length model, $\ell_{\rm turb}$, is a simplification based on the
    extraction of a viscous $\alpha$-parameter from a single
    high resolution simulation. While those effective scales should be
    correlated, they do not necessarily need to coincide, and may be different depending on the system parameters (equation of state, field topology, ...).
    Since these uncertainties both enter quadratically in the saturation value of the magnetization, they may lead to significant under- or overestimations of the saturated state.\\
    In the absence of a systematic high-resolution comparison of such
    effective models, we opt to carry out a very rough estimate of $\xi$ by
    comparing our prediction for the saturated field strength, $b^{\rm
    turb}$, due to turbulent amplification, with those observed in very high resolutions simulations \cite{Kiuchi:2015sga,Kiuchi:2017zzg,Kiuchi:2023obe}. 
    To aid this comparison, we also compute the overall equipartition field
    strength $b^{\rm eq} = \sqrt{\rho}$, which may serve as an absolute upper bound.
    The resulting outcome is shown in Fig. \ref{fig:b_turb_eq}. We can see that in the outer MRI-unstable
    layers, where the $\alpha\Omega-$dynamo is presumably active
    \cite{Kiuchi:2023obe}, the equipartition field strength reaches $b^{\rm eq}\simeq 10^{16}\,
    \rm G$. 
        \begin{figure}
    \centering
    \includegraphics[width=0.45\textwidth]{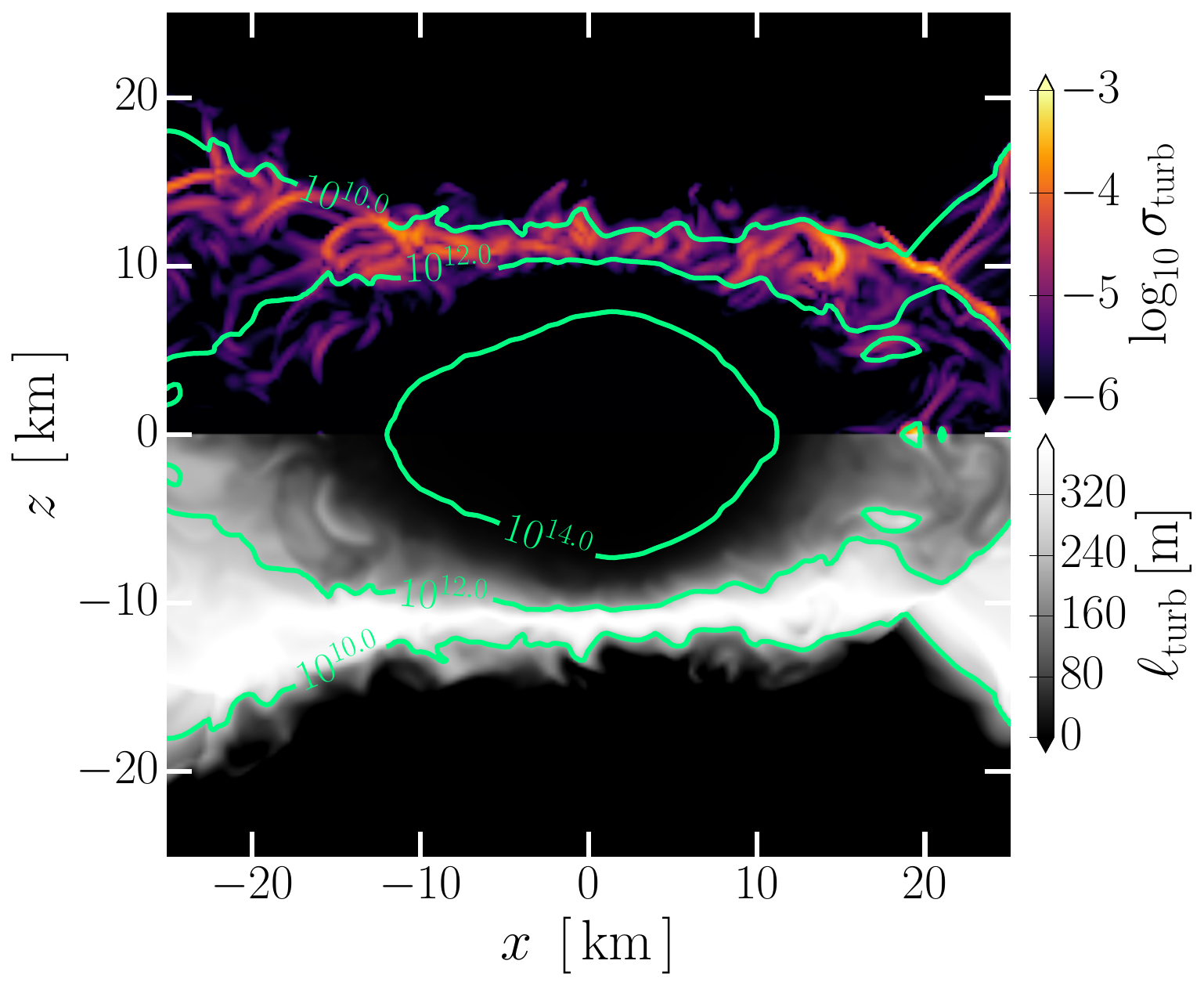}
    \caption{Turbulent mixing length, $\ell_{\rm turb}$, and corresponding
    saturation value, $\sigma_{\rm turb}$, of the magnetization $\sigma$ in
  the nascent hypermassive neutron star. The turbulent subgrid model highlights layers that are unstable to the magneto-rotational instability. The meridional plane is shown with
contour levels of the rest-mass density (in units of $\rm g/cm^3$)
indicating the structure of the star. The $\alpha\Omega-$dynamo is limited
to the outer layers of the system.}
    \label{fig:lmix}
\end{figure}
    Such high fields seem indeed to be produced in Ref. \cite{Kiuchi:2023obe} (though
    not at lower resolution simulations using large-eddy closures \cite{Aguilera-Miret:2023qih}). This is also consistent with the launch of a relativistic outflow, which requires the remnant to enter a magnetically dominated
    regime in the outer layers \cite{Most:2023sft,Combi:2023yav,Kiuchi:2023obe}. \\
    Evaluating our estimate of dynamo saturation, we can further see that for the background probed in
    the post-merger remnant, $b^{\rm turb}\lesssim 0.1 \sqrt{\xi} b^{\rm eq}$.
    This implies that the dynamo model itself likely underproduces the
    necessary magnetic field strengths, and can only jump start the self-consistent 
    evolution to higher field strengths via magnetic winding \cite{Shapiro:2000zh}. 
    This leads us to assume, that we likely
    underestimate the amount of turbulent kinetic energy from the low
    resolution simulations. At the same time, the use of $\ell_{\rm MRI}^{\rm HMNS}$ implies an implicit assumption on the resolution used to compute it. In other words, if the resolution was gradually increased the dynamo model needs to gradually switch off, meaning that $\sigma_{
    \rm turb}$ has to decrease with increasing resolution.\\
    Consequently, we heuristically introduce a dependence on the numerical grid resolution, $\Delta x$, by assuming that 
    \begin{align}
        \ell_{\rm turb} \simeq  \ell_{\rm MRI}^{\rm HMNS} \left(\frac{\Delta x}{ 12.5\, \rm m}\right)\,,
    \end{align}

    where the resolution reference scale is set by the resolution used in Ref. \cite{Kiuchi:2017zzg} from which the fit was obtained.
    Overall, this leaves us with a prescription
    \begin{align}\label{eqn:sigma_turb_final}
        \sigma_{\rm turb} = \xi \left(\ell_{\rm MRI}^{\rm HMNS}\right)^2 \left(\frac{\Delta x}{ 12.5\, \rm m}\right)^2 \sigma_{\mu\nu} \sigma^{\mu\nu}\,.
    \end{align}
    We caution that this estimate does not rely on first-principles assumptions but mainly on conclusions drawn from the few available high-resolution
    simulations in the literature. Consequently, in this work we focus on
    understanding the qualitative impact of such a dynamo prescription, rather than quantitative aspects.
    In particular, our simulations will focus on understanding the role of $\xi$ and the saturated state of the dynamo.
            \begin{figure}
    \centering
    \includegraphics[width=0.45\textwidth]{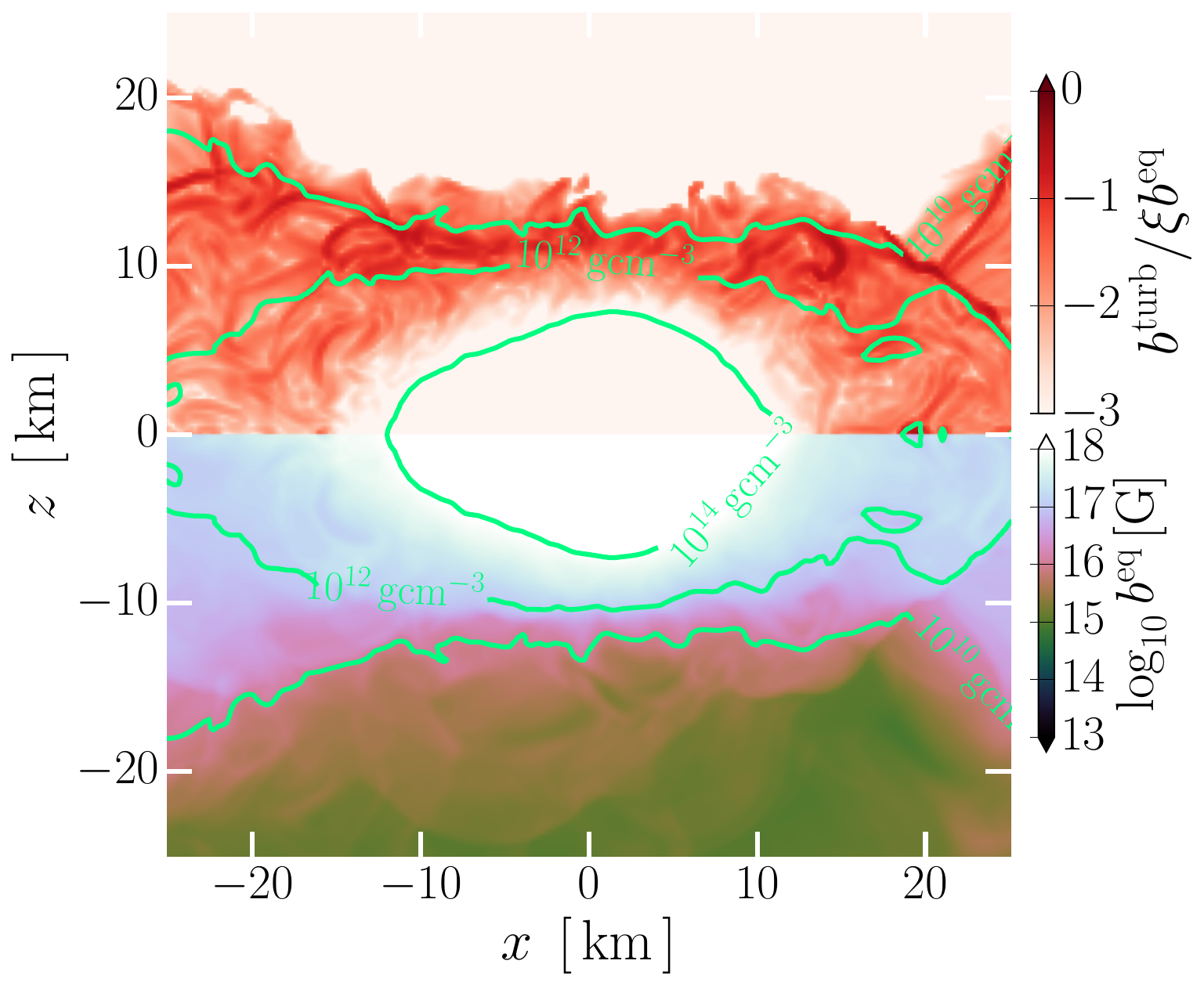}
    \caption{Estimate of the upper limit of the magnetic field strength,
      $b^{\rm turb}$, reached in equipartition with the turbulent kinetic
      energy, stated relative to the total equipartition field, $b^{\rm
      eq}\sim \sqrt{\rho}$, where $\rho$ is the rest-mass density. 
      The factor $\xi <0$ quantifies the fraction of turbulent kinetic energy converted.
      Shown is the meridional plane with
      contour levels of the rest-mass baryon density (in units of $\rm g/cm^3$)
    indicating the structure of the star.}
    \label{fig:b_turb_eq}
\end{figure}
    \item {\bf Projected impact of the model}    \\
     Finally, we estimate the impact of this model. 
     We do so by providing a back-of-the-envelope estimate for the flows present in the HMNS. 
     In general, we can estimate the characteristic magnetization limit as
    $\sigma_{\rm target} \sim
     \left(\bar{\ell}_{\rm turb}/L\right)^2 \bar v^2$, where $\bar v$ is the
     average velocity of the turbulent flow, $L$ the characteristic gradient scale,
     and $\bar \ell_{\rm turb}$ the average turbulent mixing scale.\\
     Assuming that characteristically $\bar v\sim 0.1$ inside the merger remnant, we find
    \begin{align} \label{eq:sigma_simple}
        \sigma_{\rm target}  = 0.01 \left(\frac{\ell_{\rm turb}}{L}\right)^2\,.
    \end{align}
    This simplified model for constant mixing length is what we had previously explored in Ref. \cite{Most:2023sft}, demonstrating that this will lead to postmerger flaring and the launch of a steady-state relativistic outflow.
    
\end{enumerate}
    \begin{figure}
    \centering
    \includegraphics[width=0.45\textwidth]{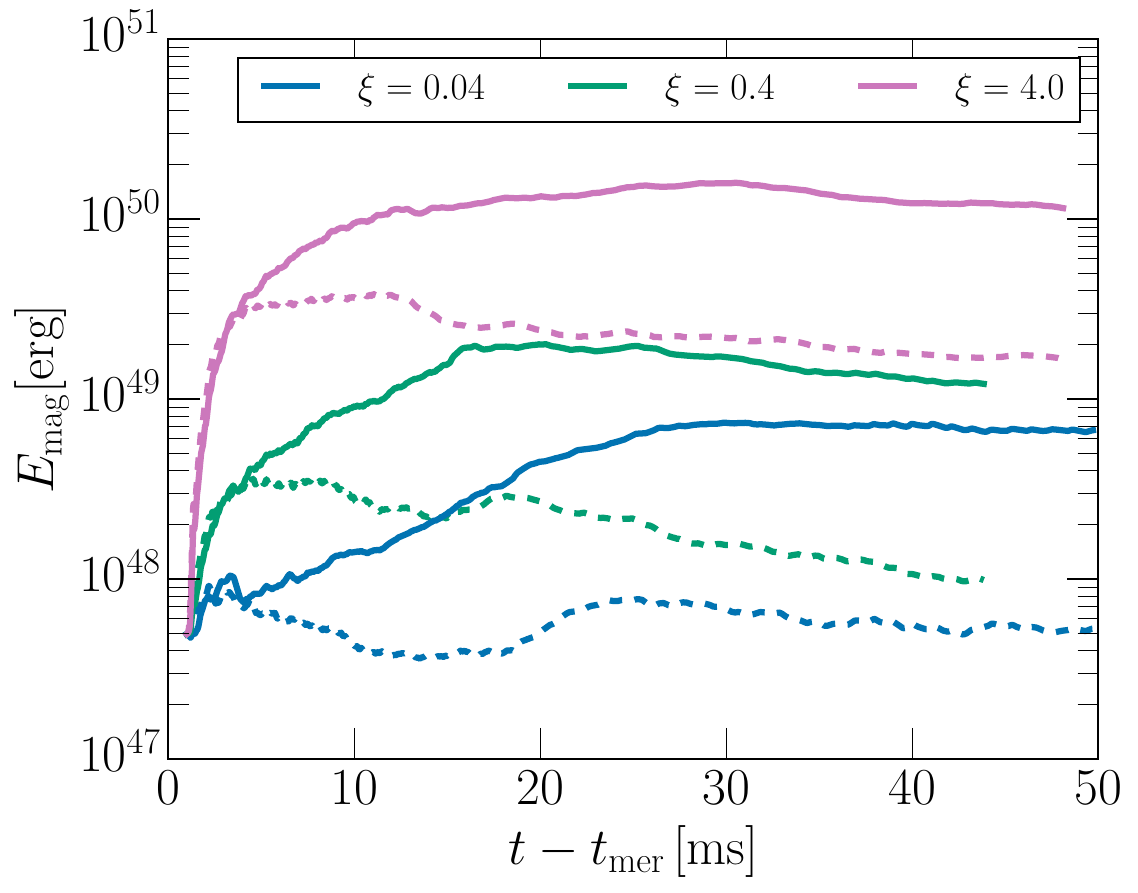}
    \caption{Average magnetic energies, $E_{\rm mag}$, for different dynamo efficiencies. Shown are toroidal (solid lines) and poloidal components (dashed lines). The poloidal component is almost always subdominant, consistent with the assumptions of the subgrid model. Different colors denote different dynamo parameters $\xi$. Times, $t$, are stated relative to the merger time $t_{\rm mer}$.}
    \label{fig:Emag}
\end{figure}
\subsection{Numerical implementation}\label{sec:numerics}
We now want to provide a brief description of how to implement 
the dynamo-augmented GRMHD equations into a numerical code. We will focus mainly on differences from standard GRMHD approaches and refer to the literature for
further details on numerical implementations of ideal GRMHD (e.g., Ref.
\cite{Duez:2005sf}).

In our implementation, we solve the Maxwell equations by evolving
a covariant vector potential $\mathcal{A}_\mu = \Phi n_\mu + A_\mu$, such
that $A_\mu n^\mu =0$, and $\Phi$ is the scalar potential
\cite{Etienne:2010ui}. Adopting Lorenz gauge, $\nabla_\mu \mathcal{A}^\mu
=0$, the evolution equations for the Maxwell sector are then given by \cite{Etienne:2011re},
\begin{align}
  \partial_t A_i &= \frac{\alpha}{u^0} \varepsilon_{ijk}
  u^j {B}^k - \alpha \kappa\left[ \left(1-v^2\right) {B}_i + \left(v_l
  B^l\right) v_i \right] \nonumber \\
  & - \partial_i \left(\alpha \Phi - \beta^j A_j \right)\,.
  \label{eqn:A_evol}
\end{align}

We solve the discrete form of Eq. \eqref{eqn:A_evol} on a staggered mesh
\cite{Etienne:2010ui}, where the electric and magnetic fields are treated using
high-resolution shock capture methods \cite{DelZanna:2002rv}. More precisely, we adopt the
upwind constraint transport method of Ref. \cite{Londrillo:2003qi}, as implemented in the ECHO
scheme \cite{DelZanna:2007pk}. An extension to non-ideal electric fields is straightforward \cite{Mignone:2019ebw}. 
No other modifications of the Maxwell equations are necessary to incorporate the scalar dynamo
term, besides Eq. \eqref{eqn:b_comov_kappa}. On the hydrodynamics side, the rescaling of the magnetic field needs to be carried out following
Eqs. \eqref{eqn:Tmunu_ideal} and \eqref{eqn:Efield_scalar}.\\
\begin{figure}
    \centering
    \includegraphics[width=0.45\textwidth]{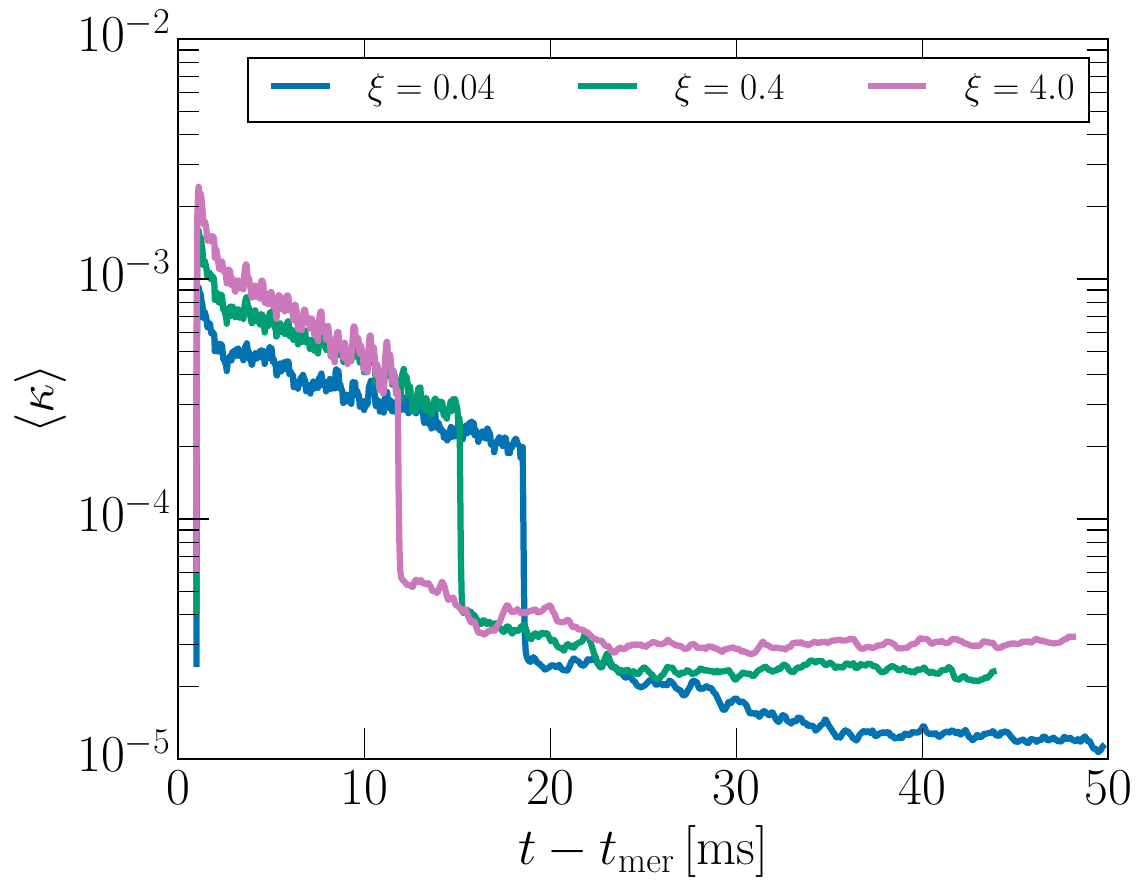}
    \caption{Density weighted average of the dynamo scalar $\kappa$ as a function of the subgrid parameter $\xi$. Once the target magnetization is reached the dynamo action switches off in most parts of the star and its disk.}
    \label{fig:kappa_avg}
\end{figure}
One important aspect of the GRMHD equations in flux-divergence form is the need to recover primitive variables
from a conserved state \cite{Noble:2005gf}.
To use standard algorithms for primitive recovery \cite{Kastaun:2020uxr}, it is necessary to decouple the direct dependence of the dynamo coefficient $\kappa$ onto the fluid state. To this end, $\kappa$, cannot directly depend on the primitive
variables $\left( \rho_b, \epsilon, u_\mu, b_\mu \right)$. Rather than 
specifying a closure relation for $\kappa = \kappa\left( \rho_b,
b_\mu\,, \ldots \right)$, we artificially introduce a
rate equation, enforcing that $\kappa$ relaxes to the desired closure
relation over a timescale $\tau_\kappa$,
\begin{align}
  \nabla_\mu \left(\kappa \rho u^\mu\right) = \frac{\rho}{\tau_\kappa}
  \left[ \kappa\left( \rho, b_\mu, \ldots \right) - \kappa \right]\,,
  \label{eqn:tildeKappa}
\end{align}
which corresponds to a simple advection equation with a relaxation term. 
The relaxation time $\tau_{\kappa}$ is a free parameter, which for simplicity we will
choose to be a constant.\\
\begin{figure*}
    \centering
    \includegraphics[width=\textwidth]{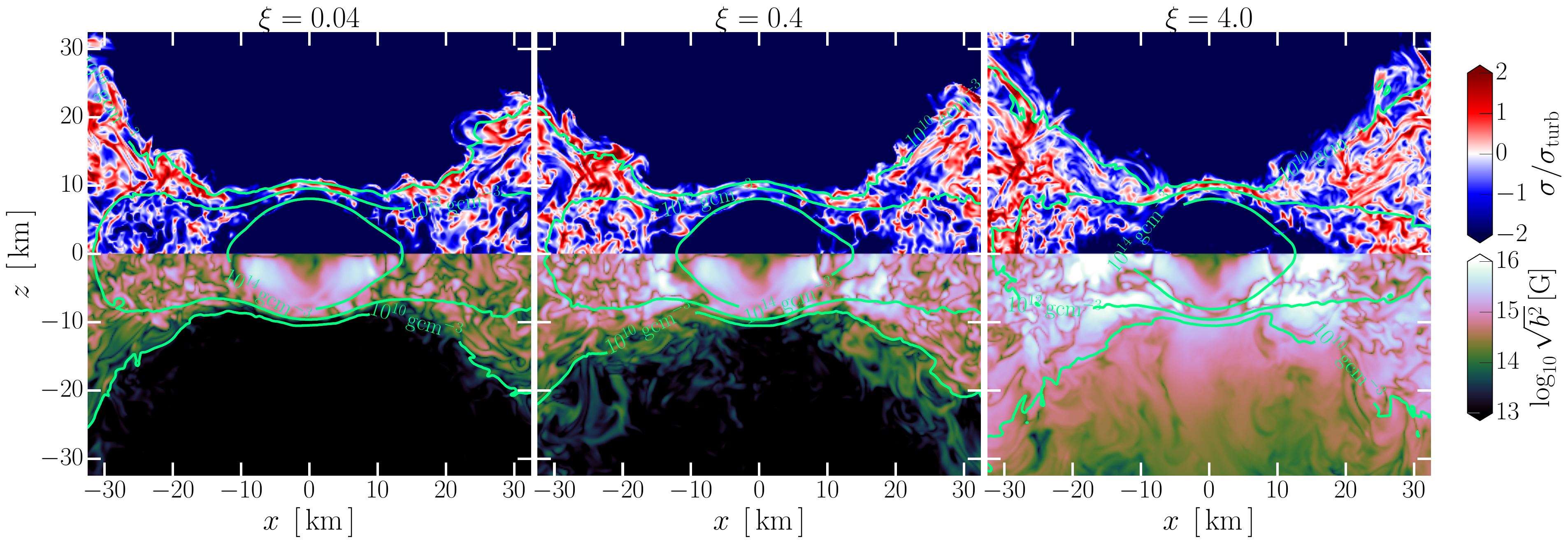}
    \caption{Saturated magnetic field inside the neutron star remnant.
    Shown are the magnetization $\sigma$ relative to the dynamo saturation scale $\sigma_{\rm turb}$, and the local comoving magnetic field strength $b$. Green level contours denote baryon density.
  The three panels correspond to different dynamo parameters $\xi$.}
    \label{fig:saturation2D}
\end{figure*}

We stress that due to the approximate nature of the
dynamo closure relation the small time lag in relaxing to the desired value
will likely have only a negligible effect on the simulation. 
At the same time, the ability to use extremely robust primitive recovery
routines only available for the ideal GRMHD system (see Ref. \cite{Ripperda:2019lsi} for challenges in
resistive GRMHD versions) is crucial to successfully performing
simulations of strongly magnetized flows when nuclear equations of state are used \cite{Kalinani:2021ofv}.

\subsection{Simulation setup}

We solve the discrete form of the magnetohydrodynamics sector 
using the \texttt{Frankfurt/IllinoisGRMHD (FIL)} code \cite{Most:2019kfe,Etienne:2015cea}.
\texttt{FIL} utilizes the ECHO scheme \cite{DelZanna:2007pk}, implemented with a fourth-order accurate
flux correction, fifth-order WENO-Z reconstruction \cite{Borges2008}, and HLLE
Riemann solver \cite{harten1983upstream}. 
Unlike previous GRMHD simulations with the code \cite{Most:2019kfe,Most:2021ytn,Chabanov:2022twz}, we have upgraded the
conservative-to-primitive inversion following the method of
\citet{Kastaun:2020uxr}. This method increases the overall accuracy of the code in
highly magnetized regions (see also \cite{Kalinani:2021ofv}), and has allowed us to study
flaring and jet launching from neutron star merger remnants \cite{Most:2023sft}. 
\texttt{FIL} also implements its own fourth-order accurate dynamical
spacetime solver using the Z4c formulation \cite{Bernuzzi:2009ex,Hilditch:2012fp}. 
\texttt{FIL} operates on top of the \texttt{EinsteinToolkit} framework \cite{Loffler:2011ay} and utilizes its
moving-box mesh-refinement infrastructure \texttt{Carpet} \cite{Schnetter:2003rb}.

The initial data for the simulation is computed using the \texttt{FUKA} code \cite{Papenfort:2021hod}, which
works on top of the \texttt{Kadath} framework \cite{Grandclement:2009ju}. 
Differently from our previous work \cite{Most:2023sft}, we consider a system with a long-lived remnant.
We adopt the DD2 equation of state \cite{Hempel:2009mc} with a total binary mass of $2.5\, \rm
M_\odot$, leading to a stable remnant neutron star. In terms of the initial mass ratio, we adopt $q=0.9$.
We initialize the magnetic field inside the two stars using a dipole magnetic field, prescribing the axisymmetric component of the vector potential in each star, $A_\phi = A_b \varpi^2 \max \left(P - 0.04\, P_{\rm max} ,0\right)^2$ \cite{Etienne:2011ea}, where $\varpi$ is the cylindrical radius, and $P_{\rm max}$ is the maximum pressure inside each star. We choose $A_b$ separately for each star, such that the maximum value of the magnetic field inside each star is $B_{\rm max} \approx 10^{15}\, \rm G$. 

We set up the numerical grid as a set of seven nested grids, which extend to
an outer separation of about $6,000\, \rm km$. The finest resolution has
been chosen to be $\Delta x = 200 \, \rm m$, which increases by a factor of two
on every succinct level. In a previous work, we have demonstrated the use
of fully high-order methods leads to a better capturing of the MRI in the disk 
at lower resolutions \cite{Most:2019kfe}.  On the other hand, properties such as the
time of jet launching and the amount of self-consistent amplification may not be well
captured compared to high-resolution simulations \cite{Aguilera-Miret:2023qih}. However, 
the resolution considered here will be sufficient for a first assessment of difference subgrid prescriptions for
the $\alpha\Omega$-dynamo.

\section{Results}\label{sec:results}

\begin{figure*}
    \includegraphics[width=0.9\textwidth]{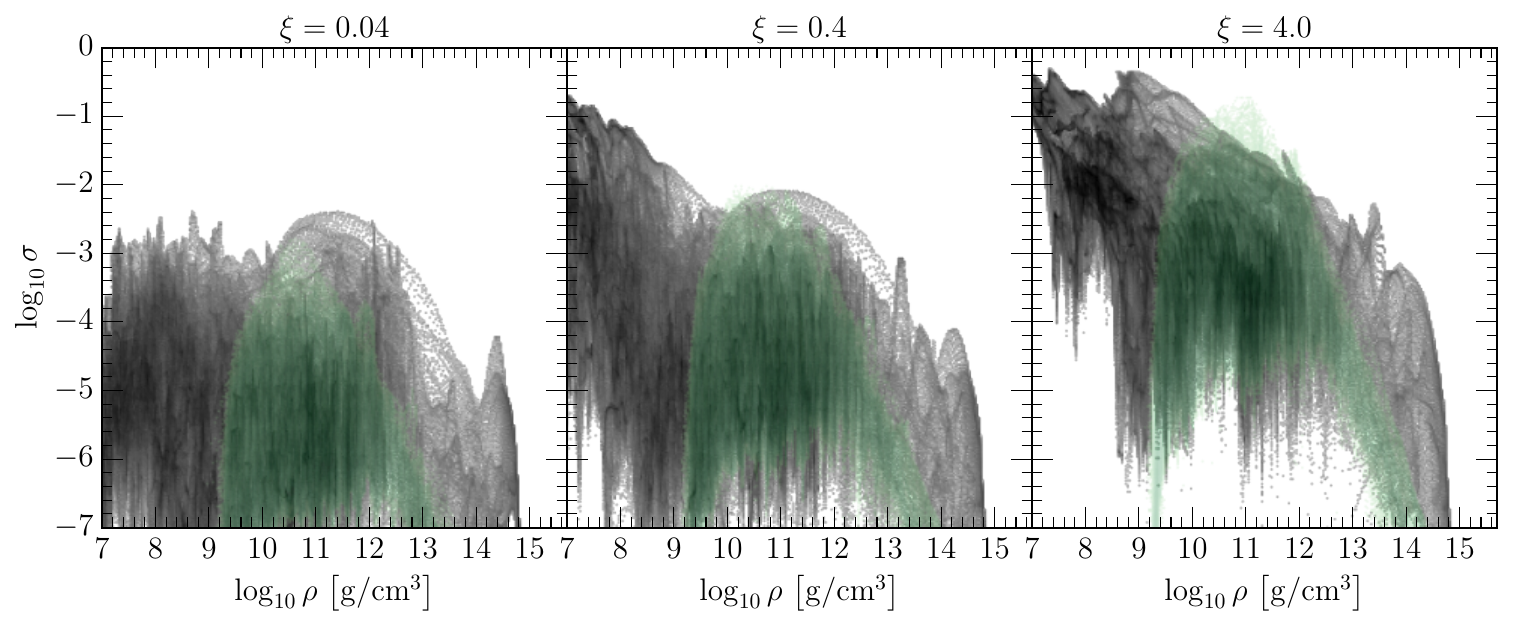}
    \caption{Magnetic energy ratio $\sigma$ for a subset of fluid elements
      in the saturated state, which are denoted by their baryon densities $\rho$.  
      Shown are the three different dynamo efficacies, $\xi$.  The green area shows the expected target magnetizations prescribed by the dynamo model.}
    \label{fig:sigma_hist}
\end{figure*}

Here we present the results of our numerical tests of different subgrid
models for the $\alpha\Omega$-dynamo. We focus on the case of a long-lived magnetar remnant,
where strong dynamo amplification could be present \cite{Kiuchi:2023obe}, aiding the launching of magnetically
driven winds and jet-like outflows from the remnant \cite{Mosta:2020hlh,Most:2023sft,Combi:2023yav}.

We summarize our discussion and motivation detailed in Sec. \ref{sec:methods} as follows:
One of the main goals of this paper is an initial investigation of the
$\alpha\Omega$- mean field dynamo model proposed in Sec. \ref{sec:methods}.
Although the growth rate we use is loosely based on recent conditions inferred from high-resolution simulations \cite{Kiuchi:2017zzg,Kiuchi:2023obe}, saturation of the dynamo cannot
be self-consistently achieved in our model and needs to be prescribed.
One of the main features of the dynamo subgrid model is that saturation is achieved once the available reservoir of turbulent kinetic energy has been
converted into magnetic energy. In the case of the MRI this will happen
only in regions of radially outward decaying differential rotation, which
only cover the outermost layers of the star \cite{Hanauske:2016gia}.
Within these regions, in principle, the total available kinetic energy
associated with the differential rotation can be converted. 
Within the turbulence-inspired subgrid model we use, the saturation uncertainty is prescribed by rescaling the target magnetization using
the parameter $\xi$. Barring the availability of a large set of
high-resolution simulations necessary for a thorough calibration of this
parameter, here we investigate three different choices, $\xi=\left(0.04, 0.4, 4.0\right)$
corresponding to different regimes produced in our standard-resolution
simulations. In this case our simulations should be regarded as complementary but not equal to recent high resolution works in the literature \cite{Aguilera-Miret:2020dhz,Aguilera-Miret:2021fre,Palenzuela:2021gdo,Kiuchi:2023obe}.

We begin by discussing various properties of the evolution of the system
associated with the dynamo subgrid model. Our discussion generally begins
in the early post-merger phase, when we activate the dynamo model. A
general description of the pre-merger and post-merger evolution of such systems can be found elsewhere (see, e.g., Refs. \cite{Baiotti:2016qnr,Radice:2020ddv} for recent reviews).

\subsection{Saturation of the $\alpha\Omega$-dynamo}

Shortly after the merger, we turn on the sub-grid dynamo model.
The dynamo will be active in the outer layers of the star and disk with the precise density dependence largely governed by Eq. \eqref{eqn:sigma_turb_final}. In the differentially rotating background flow, the $\alpha-$term will introduce an effective $\alpha\Omega$-dynamo. 
This will lead to an initial rapid growth, particularly of the poloidal magnetic field component, since the background field is largely toroidal \cite{Ciolfi:2017uak,Aguilera-Miret:2023qih}. Fig. \ref{fig:Emag} shows the evolution of the magnetic energy for all three cases considered here. 
Depending on the parameter, $\xi$, corresponding to the fraction of converted kinetic energy, the poloidal energy (dashed lines) is indeed undergoing an initial amplification phase. Afterward, the poloidal energy saturates rapidly, but the toroidal field continues to grow.
The growth is driven in the following way: In the sheared background flow, the poloidal field is wound-up leading to linear magnetic field growth due to magnetic braking \cite{Shapiro:2000zh}. At the same time, the mean field dynamo needs to sustain the poloidal field. We can more quantitatively establish this by comparing the magnetic field evolution with the duty cycle of dynamo term, as expressed in a density-weighted average $\left<\kappa\right>$ (Fig. \ref{fig:kappa_avg}). In fact, the main dynamo amplification is turned off early, and all models reach a saturated dynamo state between $10-15\, \rm ms$ after the dynamo was activated. Indeed, this is consistent with fixing the growth rate of the dynamo via a globally constant $\kappa_{\rm HMNS}$ parameter, which only switches off at saturation. We can see that coincident with the saturation of the mean field dynamo, the magnetic-field amplification slows down and at most follows a $t^2$ relation consistent with magnetic braking, or it stops entirely. The saturation levels of the magnetic energy are of the order of the fraction of converted kinetic energy, although the fractions with $\xi <1$ seem to approach a similar saturated state at late times, as a result of subsequent self-consistent magnetic field evolution.\\
\begin{figure*}
    \centering
    \includegraphics[width=\textwidth]{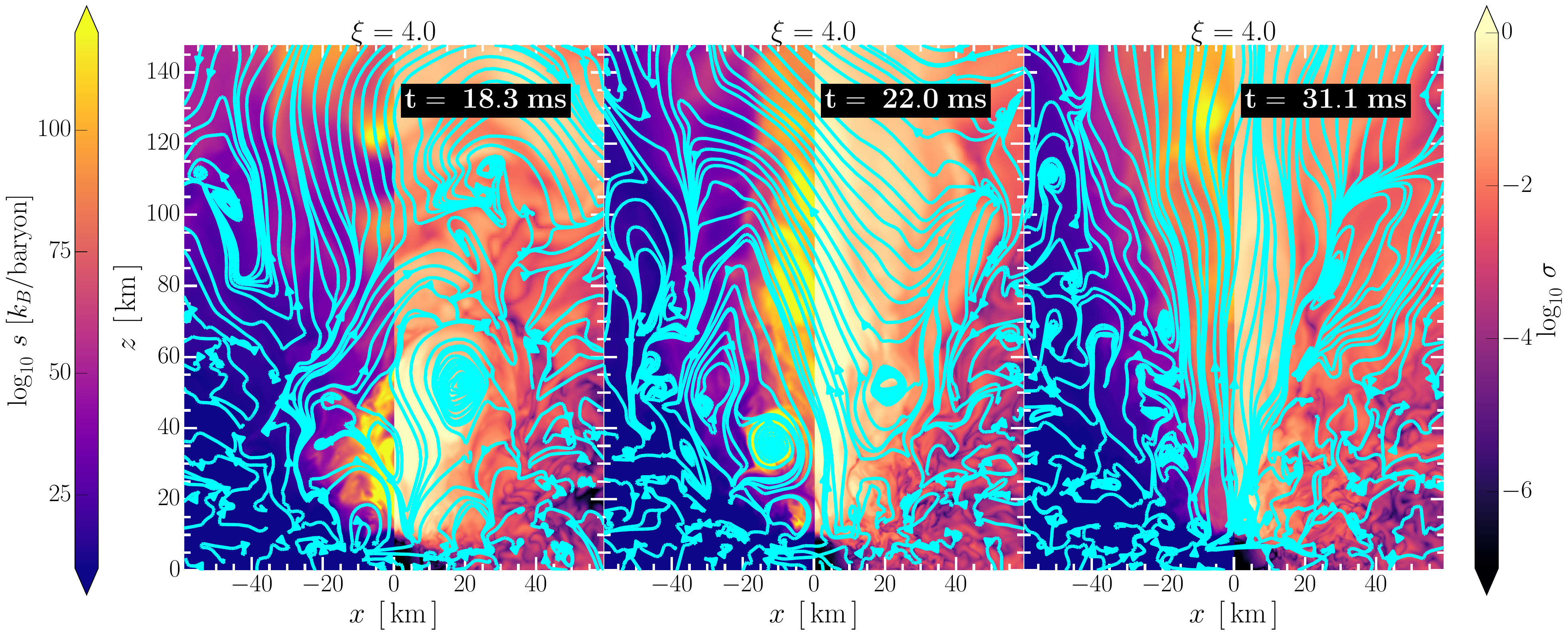}
    \caption{Initial break out and launching of a mildly relativistic outflow.
    Shown are the entropy $s$ per baryon and the magnetization $\sigma$ at different times, $t$, relative to the time of merger. Streamlines (cyan) denote the magnetic field configuration. {\it (Left)} Parker-type instabilities lead to a breakout of the magnetic field from the star. These closed loops inflate due to differential rotation, leading to a brief period of post-merger flaring activity. Shown is a post-merger flare traced out by the entropy, which is magnetically dominated. {\it (Middle)} After the flaring period, open field lines begin to rearrange in a magnetic tower geometry starting to drive a mildy relativistic outflow.
    {\it (Right)} The outflow is sustained but baryon pollution leads to a gradual reduction in magnetization of the polar region. }
    \label{fig:sigma_star}
\end{figure*}
\begin{figure}
    \centering
    \includegraphics[width=0.49\textwidth]{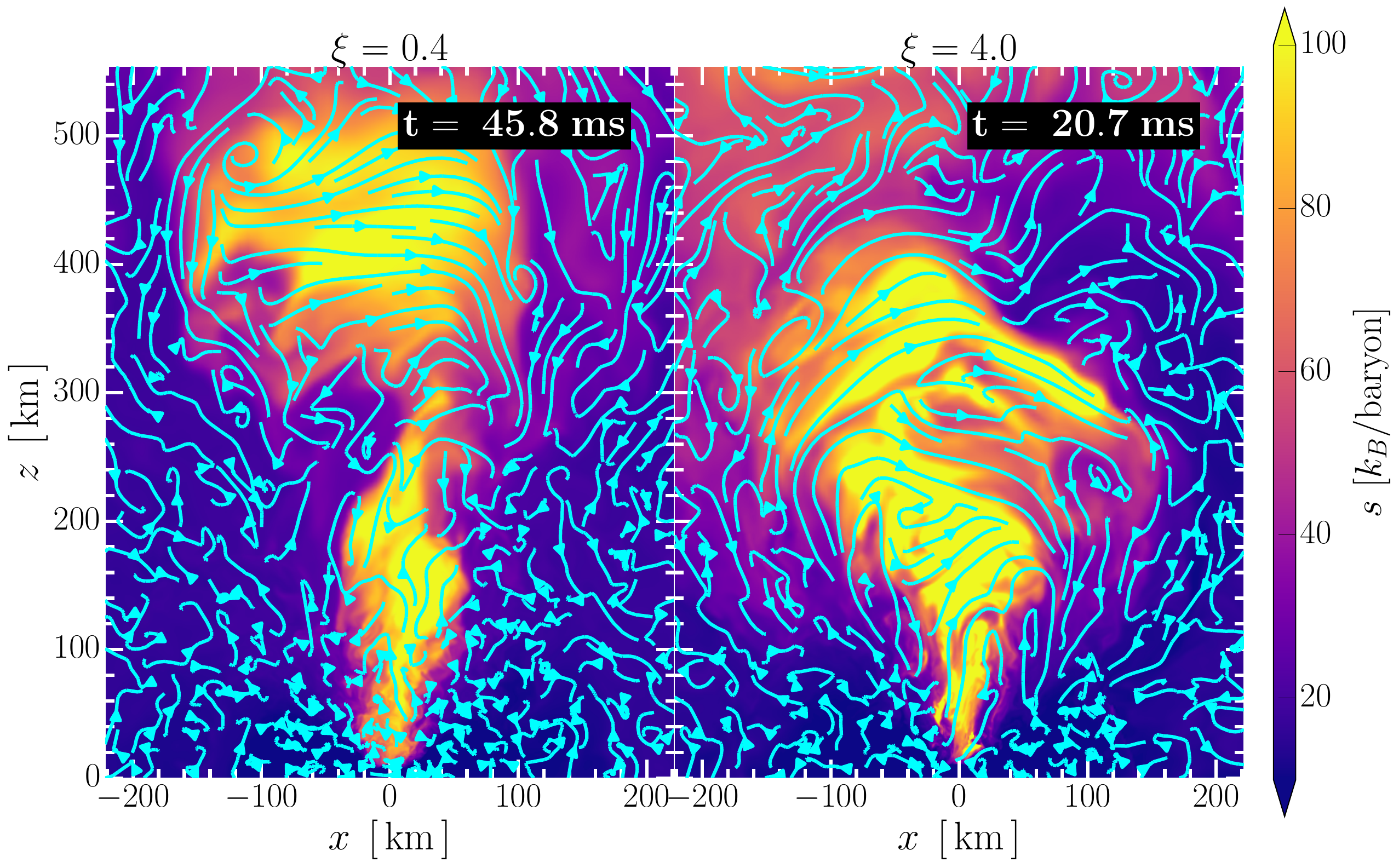}
    \caption{Flares emitted prior to launching a sustained magnetically
    driven outflow. Several flares are intermittently launched, which are
traced by using the entropy $s$ per baryon. The magnetic field structure is shown in cyan. Different times $t$ reflect the different time of
magnetic field breakout from the star. Shown are configurations for medium and high values of the dynamo amplification parameter $\xi$.}
    \label{fig:flares}
\end{figure}

We can more quantitatively understand the saturation properties by considering the
spatial distribution of the saturated magnetization inside the neutron-star remnant (Fig.
\ref{fig:saturation2D}). We can see that in the saturated state large regions close to the surface of the star and disk have undergone dynamo amplification.
Further self-consistent magnetic field evolution has then taken over and
has managed to amplify the field beyond the prescribed saturation level,
typically by about an order of magnitude. 
We find that the local field strength close to the surface of the star
can reach $b\sim 10^{15}\, \rm G$ for weak amplification, and exceed $10^{16}\,
\rm G$ for strong amplification. Within our discussion about expected
saturated field strength (Fig. \ref{fig:b_turb_eq}), these levels are higher than what the prescribed mean field dynamo can
achieve on its own, given the imposed saturation bound. 
This is fully consistent with subsequent amplification being active by
means of winding.
The major difference between the simulation with varying subgrid
parameter is the presence of substantial amplification in the disk region,
at distances $20-30\,\rm km$ from the origin. Here, in the case of weak
dynamo driving, no substantial amplification is active. In the other cases,
additional amplification is present, enhancing the magnetic field strength
in the disk. Since winding would be active in either case, it seems likely
that additional amplification has aided the growth (and numerical
capture) of the MRI in these regions (see also \cite{Siegel:2013nrw,Most:2019kfe}). 
In the strongest magnetic field case,
this leads to a clear breakout of the field not only from the stellar surface
but also from the disk region, injecting substantial fields of $10^{15}\,
\rm G$ there. We can quantify this more explicitly by comparing the
injection of field with the dynamo subgrid model to the actual
magnetizations reached in the remnant. In Fig. \ref{fig:sigma_hist} we do so, showing the distribution of the magnetization of fluid cells from the remnant as a function of
density. We can clearly see the dynamo injection range acting on densities
between $10^9-10^{13}\, \rm g/cm^3$. In the intermediate and strong
amplification case, we can see that magnetization cascades to lower
densities, driving amplifications in layers with densities $<10^9 \, \rm
g/cm^3$. To sustain amplification in those regions, significant mixing of
matter in denser layers with low density layers needs to happen. This may be associated with spiral winds injecting magnetized material into the disk \cite{Nedora:2019jhl,Radice:2023xxn}. This leads
us to speculate that dynamo amplification in intermediate layers of the
star may qualitatively affect the outcome of a neutron star merger
simulation.

\begin{figure*}
    \centering
    \includegraphics[width=\textwidth]{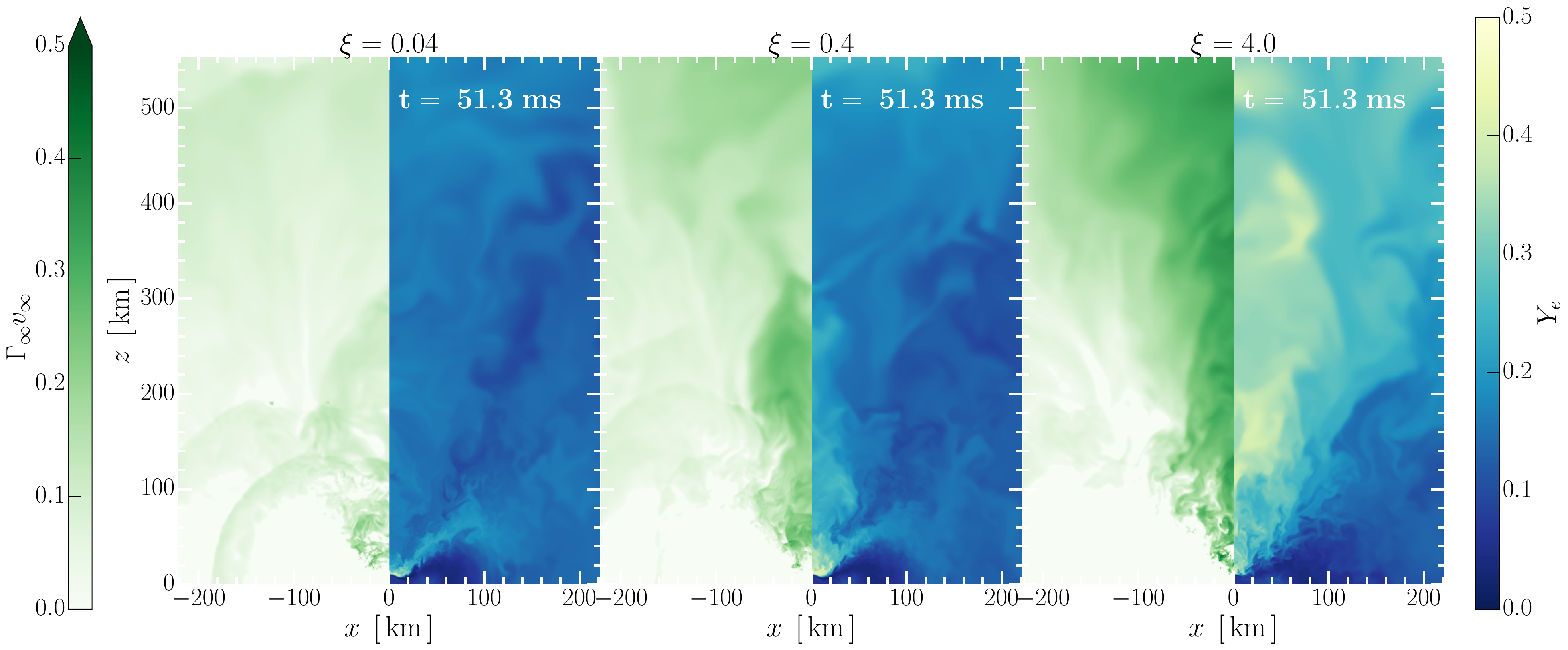}
    \caption{Large scale structure of the mildy relativistic stellar outflow. 
      Shown are terminal velocity, $\Gamma_{\infty}
      v_{\infty}$, at infinity, and the electron fraction, $Y_e$,
for all dynamo efficiencies, $\xi$.
     Times, $t$, are stated relative to the time of merger. Shown is the meridional plane for a fixed time $t$ after merger.
    }
    \label{fig:sigmaW}
\end{figure*}

\subsection{Breakout of the magnetic field and flaring emission}
Two of our models $(\xi = 0.4, 4.0)$ feature break-out of the magnetic
field and the launching of a mildly relativistic magnetically driven
outflow. In the following, we will provide a brief description of this
launching mechanism, and also describe the emission of flares which could
power precursors to short gamma-ray bursts \cite{Most:2023sft}.\\
We begin by illustrating the breakout mechanism using the
highest dynamo amplification case, $\xi=4.0$.  
After an initial amplification of the magnetic field due to the mean field
dynamo, the toroidal field gets significantly amplified due to magnetic
braking \cite{Shapiro:2000zh}. At some point, the magnetic pressure in the
toroidal field is substantial to make field lines rise magneto-bouyantly by
means of a Parker instability \cite{1978RSPTA.289..459A}. See
also Ref. \cite{Kiuchi:2011yt} and \cite{Most:2023sft} for simulations of the Parker instability in isolated and remnant neutron stars.
In Fig. \ref{fig:sigma_star}, we show that for our present simulations
the initially confined magnetic
field will magneto-bouyantly rise, emerging as closed loops connected to
different points at the surface of the star. This can most easily be seen
in terms of the entropy $s$ tracing out these magnetically dominated flux
tubes (left panel). Differential rotation of the star will further inflate
these loops. After a relative twist of about $180^\circ$, the base field
lines of the flare will reconnect, leading to the ejection of post-merger flares \cite{Most:2023sft}.
This observed mechanism is overall very similar to the reported phenomenology
in simulations of magnetar giant flares \cite{Parfrey:2013gza,Carrasco:2019aas,Mahlmann:2023ipm}, and precursor flares from
in the inspiral of a neutron star binary \cite{Most:2020ami,Most:2022ojl,Most:2023unc}.
While this process could in principle also driven by chemical gradients,
these parts of the  remnant are stable against convection consistent with
recent finidings of stable stratification inside the remnant
\cite{Radice:2023zlw}. \\
In our present simulations, we observe about three flaring events. We show
the large-scale morphology of these flares in Fig. \ref{fig:flares}.
We find that the emission of flares in both cases is intermittent,
and their strength depends on the dynamo action. We will discuss this in
more detail in Sec. \ref{sec:outflows}.

After launching of initial flares, these
are immediately followed by an intermittent magnetically dominated
outflow, leading to an initial collimation of the polar magnetic field
(middle panel, Fig. \ref{fig:sigma_star}). This collimation resembles a magnetic tower 
configuration \cite{1996MNRAS.279..389L} (see also Refs. \cite{Shibata:2011fj,Bromberg:2015wra} for a discussion in the neutron star context).
Eventually, this outflow relaxes to a steady state (see Fig. \ref{fig:Mej_t}), whose properties we discuss
further in Sec. \ref{sec:outflows}. While initially the polar outflows are
magnetically dominated, the increase in mass ejection rate leads to a
reduction of the magnetization making the outflow  at late times less
magnetized. This effect of baryon loading polar outflows is consistent with
previous simulations of proto-neutron stars \cite{Dessart:2008zd}, and will eventually
lead to a decline in the electromagnetic Poynting flux observed in this
system (Fig. \ref{fig:LEM}), see also Ref. \cite{Metzger:2018qfl}. We stress that this effect could be
further enhanced if neutrino absorption was included in our simulations
\cite{Curtis:2023zfo}.  \\

\subsection{Magnetically driven winds and outflows} \label{sec:outflows}

Shortly after the initial break-out of the magnetic field from the magnetar
remnant, a quasi-steady magnetically driven polar outflow begins to
develop. Such an outflow may be crucial in order to explain a neutron-poor
component of the kilonova afterglow \cite{Metzger:2018qfl,Mosta:2020hlh,Combi:2023yav,Curtis:2023zfo}. In addition, the outflow may
carry substantial electromagnetic (Poynting) flux, making it potentially
relevant in the context of short gamma-ray burst production \cite{Combi:2023yav,Most:2023sft,Kiuchi:2023obe}.\\
In the following, we want to provide a brief description of the polar outflow
properties and their dependence on the dynamo parameter.\\
\begin{figure*}
    \centering
    \includegraphics[width=0.9\textwidth]{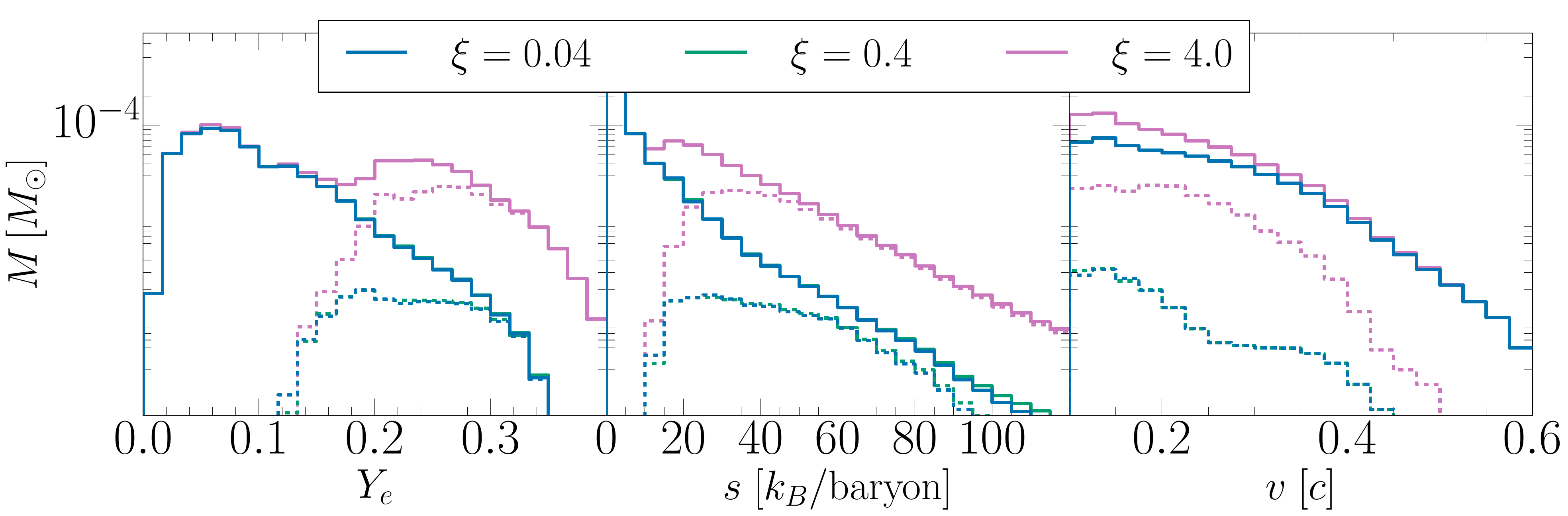}
    \caption{Histograms of mass ejection properties. Shown are the ejected mass
    $M$ as a function of electron fraction $Y_e$, entropy $s$ per
  baryon and velcity $c$. Colors denote different dynamo parameters $\xi$,
  solid lines the total ejecta, dashed lines only polar ejecta ($\theta <
30^\circ$).}
    \label{fig:ejecta}
\end{figure*}

\begin{figure}
    \centering
    \includegraphics[width=0.45\textwidth]{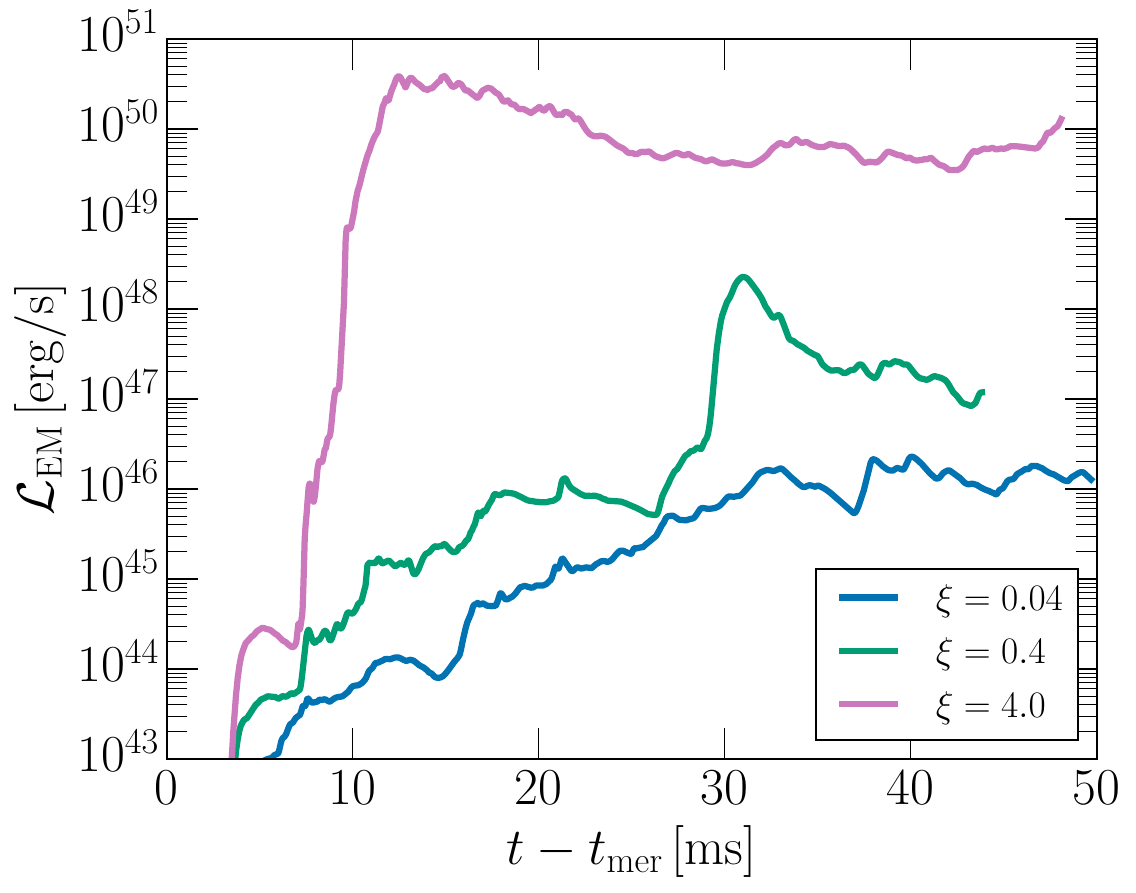}
    \caption{Electromagnetic (Poynting) flux, $\mathcal{L}_{\rm EM}$,
  measured at $r= 236\, \rm km$ from the remnant. Different curves
correspond to the different dynamo parameters $\xi$. All times $t$ are
stated relative to the merger time $t_{\rm mer}$. }
    \label{fig:LEM}
\end{figure}

We begin by discussing the overall baryon loading and energetic properties
of the outflow, such as the terminal Lorentz factor,
$\Gamma_\infty$(Fig. \ref{fig:sigmaW}).
We can estimate the terminal velocity $\Gamma_\infty v_\infty$ 
using the conservation of Bernoulli's constant $h u_t$, where $h$ is
the specific enthalpy, which serves as an upper bound on $u_t \simeq \Gamma$ if fully converted into a kinetic energy at large distances. Overall, we therefore set
$\Gamma_\infty \simeq - h u_t$, and observe that the outflow will likely
only reach mildly relativistic velocities $\Gamma_\infty v_\infty \simeq
0.5$. We caution that this picture may in principle get modified by
additional neutrino energy deposition in the ejecta, e.g., due to pair
annihilation \cite{Narayan:1992iy,Just:2015dba,Foucart:2016rxm,Fujibayashi:2017xsz}. The full details of this process, as well as the
interplay with the expected concurrent increase in baryon loading of the
outflow will sensitively depend on the method to model neutrino radiation
\cite{Foucart:2018gis,Curtis:2023zfo}. 
Within the approximations made in this work, we can also clearly
see that the Lorentz factor as well as the amount of collimation of the
outflow depends sensitively on the employed dynamo amplification. For the
weakest dynamo amplification case ($\xi = 0.04$) we observe no collimated
polar outflow and only very low density winds with velocities $v_\infty
\lesssim 0.1$. For medium amplification ($\xi = 0.4$), we do find a
collimated outflow after $t \simeq 40\, \rm ms$ with velocities $v_\infty
\gtrsim 0.2$. In the case of the highest amplification ($\xi = 4.0$), we do
observe the strongest outflow reaching velocities of $v_\infty = 0.4$.
Comparing this with the magnetization, we realize that these polar winds
are always strongly baryon loaded, i.e., have $\sigma <1$. This is
consistent with the reduction of the terminal ejecta velocity \cite{Dessart:2008zd}, however, the details may strongly depend on the amount of dynamo amplification and could probably be higher than what we investigate here \cite{Combi:2023yav,Kiuchi:2023obe}.\\
That said, if we in addition consider the Poynting flux from the system (Fig.
\ref{fig:LEM}), we do
find that the highest amplification case ($\xi = 4.0$) does reach
substantial electromagnetic (Poynting) fluxes, $\mathcal{L}_{\rm EM}$,
reaching a quasi-steady state in excess of $\mathcal{L}_{\rm EM} \simeq 10^{50}\, \rm erg/s$.
Such a value is below those found by Ref. \cite{Kiuchi:2023obe} using ab-initio high-resolution simulations of this process, which claim the strong presence of
an $\alpha\Omega-$dynamo. For less efficient amplification values, we find
that the luminosity is roughly consistent with $\mathcal{L}_{\rm EM}
< 10^{47}\, \rm erg/s$, even when a collimated outflow is present.
We caution that for larger value of $\xi$ we may likely observe a larger amplification, although the energy budget of the remnant in the outer layers would likely need to be modified compared to what our standard resolution simulations are able to produce.

\begin{figure}
    \centering
    \includegraphics[width=0.45\textwidth]{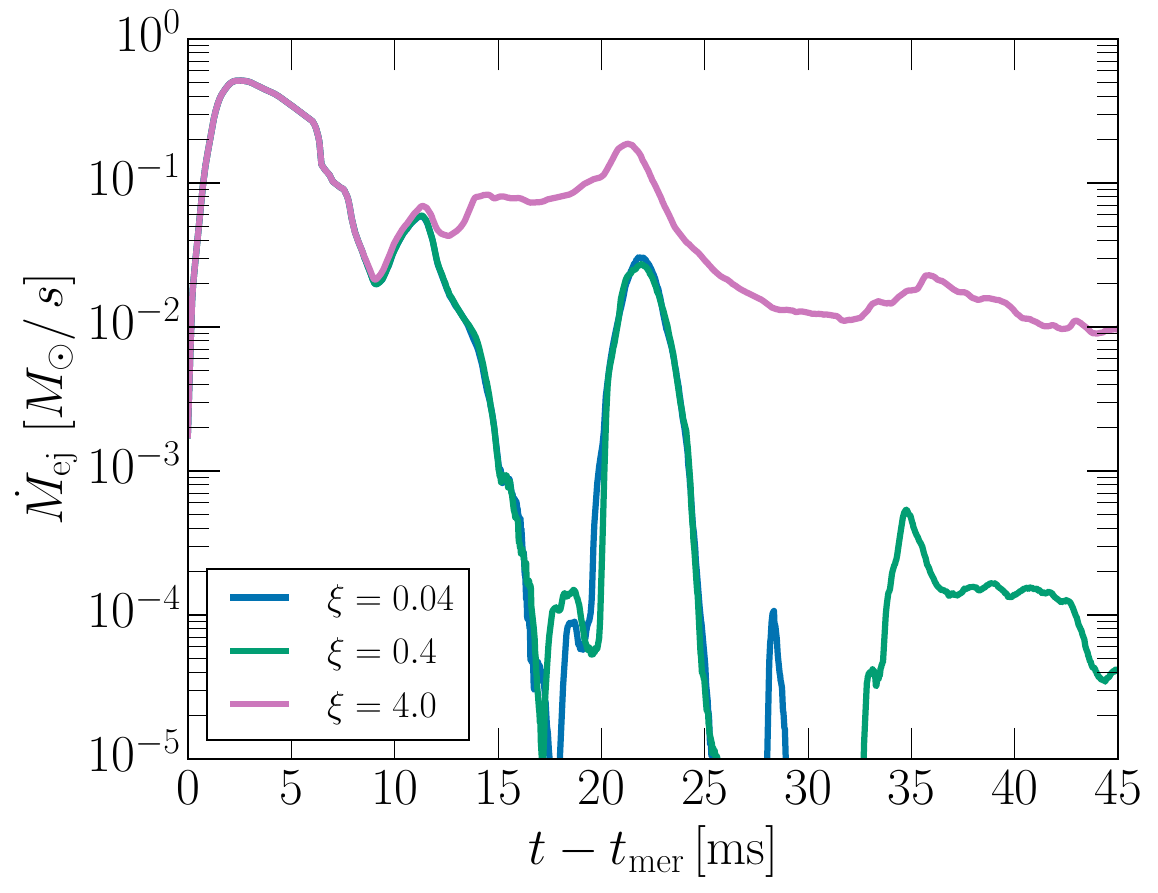}
    \caption{Mass ejection flux $\dot{M}_{\rm ej}$. Launching of magnetically-driven polar outflows leads to a
    small increase in ejection, and quasi-steady flux at late times, $t$,
    relative to the time of merger $t_{\rm mer}$. 
    Different colors correspond to different dynamo parameters $\xi$.}
    \label{fig:Mej_t}
\end{figure}

We can also quantify the nuclear composition of the outflow. We do so by
tracking the electron fraction at large distances. Previous numerical studies have
suggested that magnetically driven polar outflows may
be required to provide additional protons necessary to match
observed kilonova afterglows \cite{Mosta:2020hlh,Combi:2023yav,Curtis:2023zfo}, (but see also, e.g., Ref. \cite{Miller:2019dpt}, for
disk models). In Fig. \ref{fig:sigmaW}, we
depict the nuclear composition in terms of the proton fraction $Y_e$.
We can see that the low magnetization case ($\xi=0.04$) does not feature any enhancement of $Y_e$. On the other hand the cases that do feature break-out of the field drive midly relativistic outflows, and especially the simulation with $\xi=4.0$ features substantial proton-rich winds coming from the star, albeit with only very mildly relativistic velocities.\\
We can also provide a more quantitative assessment of the properties of these
outflows. In Fig. \ref{fig:ejecta}, we show the compositional and entropy
properties of the flows. We confirm the presence of a second
peak around $Y_e\sim 0.25$ in addition to the bulk of the dynamical ejecta,
which are very neutron rich $Y_e < 0.1$. Similarly, the ejecta have
significantly higher entropy when the dynamo action leads to the launching
of mildly relativistic mass ejection. These results are roughly consistent
with previous studies that only used neutrino cooling \cite{Mosta:2020hlh}, but are lower
than what would be expected for polar outflows when neutrino absorption was
included \cite{Dessart:2008zd,Curtis:2023zfo}.
We can finally comment on the mass flux of the ejecta $\dot{M}_{\rm ej}$ ( Fig. \ref{fig:Mej_t}).
After the dynamical ejecta have propagated out ($t> t_{\rm mer} + 30\, \rm
ms$), we can see the presence of a sustained outflow for dynamo parameters
supporting magnetically driven polar outflows. In the case of strong
amplification ($\xi=4.0$), we find a sustained quasi-steady outflow rate, which
is slightly lower than the dynamical mass ejection rate, $\dot{M}_{\rm ej}
\approx 10^{-2} (10^{-4}) M_\odot/\rm s$ for the dynamo parameters $\xi = 4.0
(0.4)$.  This steady-state outflow rate is in good agreement with recent
simulations in full numerical relativity \cite{Combi:2023yav}. Our results then indicate
that -- if these rates were sustained -- strong dynamo amplification in the
surface layer of the star may be needed to drive sustained outflows.
It is important to mention that the main purpose of these simulations is to perform a first evaluation of the $\alpha\Omega-$ subgrid dynamo model
presented in Sec. \ref{sec:methods}. We therefore consider only short
simulations $t\lesssim\, 50 \rm ms$, over which secular mass ejection driven
largely by the disk has not yet fully set in. However, secular mass
ejection is likely to dominate the total amount of ejecta \cite{Radice:2020ddv}. 
In the case of a purely viscously driven evolution, mass ejection from the star 
has been shown to be variable at late times \cite{Fujibayashi:2017puw,Fahlman:2023qig}.

\section{Conclusions}

We have investigated the impact of an $\alpha\Omega$-dynamo in a magnetar remnant formed
in a binary neutron star merger. This was done using a new approach to the $\alpha-$dynamo mean field equations
in GRMHD, which resembles the Newtonian formulation of the equations.
Based on heuristically motivated closure relations for an MRI-unstable layer in the HMNS, we have performed several simulations varying the degree of dynamo saturation in the star.
We find that for strong amplification in excess of a magnetization $\sigma \gtrsim 10^{-3}$,  magnetic winding will generate strong toroidal fields, leading to an eventual magneto-bouyant break out of the field lines from the star by means of a Parker instability \cite{1978RSPTA.289..459A,Kiuchi:2017zzg}. This is consistent with previous work \cite{Most:2023sft,Combi:2023yav,Kiuchi:2023obe} showing that these field lines will then rearrange into a magnetic tower configuration \cite{1996MNRAS.279..389L}, which may power intermittent outflows \cite{Kiuchi:2023obe}. We find that in the case of sufficient dynamo amplification, these will be preceded by a short period of flares \cite{Most:2023sft}, which may be able to power a short gamma-ray burst precursor \cite{Gottlieb:2023sja}. Crucially, we find that for strong amplification the Poynting flux can reach $\mathcal{L}_{\rm} \simeq 10^{50}\, \rm erg/s$, while for a weaker dynamo strength it will never exceed $10^{47}\,\rm erg/s$. Higher dynamo amplification than considered in this work would likely drive stronger Poynting fluxes. Similarly, we find that the terminal Lorentz factor of the magnetically driven winds as well as the degree of proton-richness of the outflows correlate strongly with the dynamo amplification. We caution that the results may be strongly affected by the degree of baryon loading, which in our simulations will be underestimated due to the lack of neutrino absorption in the simulation \cite{Dessart:2008zd,Foucart:2018gis}. Clarifying this point will require simulations using full neutrino transport of the postmerger remnant \cite{Radice:2023zlw,Curtis:2023zfo}.\\
Overall, the picture we find for strong dynamo amplification, however, appears qualitatively consistent with recent works in the literature \cite{Mosta:2020hlh,Combi:2023yav,Curtis:2023zfo}, even though our electromagnetic fluxes appear to be lower. To fully determine the realistic amount of dynamo amplification as well as performing a calibration of the subgrid model proposed here will require high-resolution simulations able to self-consistently capture the precise amount of amplification present in the system \cite{Kiuchi:2023obe,Chabanov:2022twz,Aguilera-Miret:2023qih}. This leads us to conclude in line with recent very high-resolution simulations of the $\alpha\Omega-$dynamo, that strong magnetic field amplification in the outer layers of the star may be an important ingredient for our understanding of whether magnetar remnants can power gamma-ray bursts.

\section*{Acknowledgments}
The author gratefully acknowledges insightful discussions with A. Beloborodov, A. Bhattacharjee, L. Combi, P. M\"osta, C. Musolino, L. Rezzolla, and E. Quataert. 
This work is supported by the National Science Foundation under grant No.
PHY-2309210.  This work mainly used Delta at the National Center for
Supercomputing Applications (NCSA) through allocation PHY210074 from the
Advanced Cyberinfrastructure Coordination Ecosystem: Services \& Support
(ACCESS) program, which is supported by National Science Foundation grants
\#2138259, \#2138286, \#2138307, \#2137603, and \#2138296.  Additional
simulations were performed on the NSF Frontera supercomputer under grant
AST21006.  
We acknowledge the use of the following software packages: EinsteinToolkit
\cite{Loffler:2011ay}, FUKA \cite{Papenfort:2021hod}, Kadath \cite{Grandclement:2009ju}, kuibit \cite{kuibit}, matplotlib
\cite{Hunter:2007}, numpy \cite{harris2020array}, scipy
\cite{2020SciPy-NMeth}. 

\appendix

\section{Turbulent kinetic energy estimate}\label{app:turb}
Here we want to provide some details on the turbulent kinetic energy saturation criterion used in Sec. \ref{sec:subgrid}.
Specifically, we adopt the approach of Refs. \cite{2011A&A...528A.106S,2022MNRAS.513.6028L,Miravet-Tenes:2022ebf,Miravet-Tenes:2023see} to model a proxy for the turbulent kinetic energy 
in the implicit large eddy paradigm, i.e., without including explicit viscous and resistive scales. 
Rather than adopting a full closure model \cite{Carrasco:2019uzl,Vigano:2020ouc}, our aim is to provide a simple way of incorporating a mean field dynamo effect, which does not compromise the stability of current GRMHD codes \cite{Kastaun:2020uxr,Kalinani:2021ofv}.
As a starting point, we assume that the turbulent energy in the implicit large-eddy approach can be related to the square of the shear tensor, $\sigma_{\mu\nu}$ \cite{2011A&A...528A.106S}. 
Using the formalism of dissipative hydrodynamics \cite{denicol2022microscopic}, one can write the entropy current for a flow exhibiting shear stresses and resistive dissipation, as 
\begin{align}\label{eqn:IS}
    \nabla_\mu \left(\rho s u^\mu\right) = \frac{2 \rho \nu}{T}\sigma_{\alpha \beta} \sigma^{\alpha \beta} + \frac{1}{T} j^\kappa e_\kappa\,,
\end{align}
where $T$ is the fluid temperature, and the terms on the right correspond to viscous and resistive heating, respectively. Here, $j^\kappa$ are the electric current and $\sigma^{\alpha\beta}$ is the covariant shear tensor. 
While viscous stresses vanish for a perfect fluid, we can easily clarify that the dynamo model proposed here does not contribute to entropy production at the expansion order considered.  For simplicity we assume $e^\mu \sim \eta j^\mu$, 
where $\eta$ is the effective (numerical resisitvity). Since then $j^\kappa e_\kappa \sim \kappa^2 \eta b^2$,the entropy density $s$ is conserved up to the validity of our dynamo approximation,
\begin{align}
    \nabla_\mu \left(\rho s u^\mu\right) = 0 + \mathcal{O}\left(\kappa^2\right)\,.
\end{align}
In line with Refs. \cite{2011A&A...528A.106S,2022MNRAS.513.6028L}, we now need to identify an effective turbulent kinetic energy, $\varepsilon_{\rm turb}$ associated with the turbulent kinetic energy budget available. Due to the similarities of the Newtonian equations with Eq. \eqref{eqn:IS}, we heuristically propose
\begin{align}
    \nabla_\mu \left(\frac{\rho \varepsilon_{\rm turb}}{T} u^\mu\right) = \frac{2 \rho \nu_{\rm turb} }{T}\sigma_{\alpha \beta} \sigma^{\alpha \beta} - \frac{1}{T\tau_{\rm turb}} \rho \varepsilon_{\rm turb} +\mathcal{O}\left(\kappa^2\right)\,.
\end{align}
Note that $\varepsilon_{\rm turb}$ is not conserved and decays on a timescale, $\tau_{\rm turb}$, consistent with the effective turbulent energy cascade. In turn, $\tau_{\rm turb}$ is a new free parameter of this system.\\

In practice, on macroscopic (slow) timescales the system will likely approach a steady state equilibrium, between driving and decay of turbulence.
In this case, the system relaxes such that $\nabla_\mu \left(\rho \varepsilon_{\rm turb}/T \right) \simeq 0 $. This leads to a steady state turbulent kinetic energy
\begin{align}
    \varepsilon_{\rm turb} = 2 \nu_{\rm turb} \sigma_{\alpha \beta} \tau_{\rm turb} \sigma^{\alpha \beta}\,.
\end{align}
For the application considered here, we can correlate the turbulent viscosity $\nu_{\rm turb}$ with a mixing length, $\ell_{\rm turb}$ \cite{Radice:2017zta,Radice:2020ids}.
On dimensional grounds we may write $\nu_{\rm turb} = \ell_{\rm turb} c_{\rm flow}$, where $c_{\rm flow} \simeq c_s$ is the characteristic speed of the medium corresponding to the sound speed, $c_s$, in hydrodynamic turbulence, or to the Alfven speed $v_A$ for Alfvenic turbulence.
Due to the steady state approximation, we further make the assumption that the turbulent decay timescale $\tau_{\rm turb} \simeq \ell_{\rm turb} / c_{\rm flow}$. 

Following Ref. \cite{2022MNRAS.513.6028L}, we now assume that the turbulently driven dynamo action is quenched, when the magnetic energy density $\varepsilon_{\rm mag}= \frac{1}{2} b^2$ reaches a critical fraction $\xi_{\rm turb} <1$ of the turbulent energy budget, $\rho \varepsilon_{\rm turb}$. Put differently, saturation should set in when a target magnetization $\sigma_{\rm turb} = 2 \xi_{\rm turb} \varepsilon_{\rm turb}$ is reached. 
In summary, this leads to an effective magnetization set by turbulence as
\begin{align}
    \sigma_{\rm turb} = \xi \ell_{\rm turb}^2 \sigma_{\alpha \beta} \sigma^{\alpha \beta}\,,
\end{align}
where for convenience we have absorbed the remaining constant numerical prefactor into $\xi$.

\bibliography{inspire,non_inspire}
\end{document}